\documentclass{article}
\usepackage[affil-it]{authblk}
\usepackage[utf8]{inputenc}
\usepackage{geometry}
\usepackage{caption,subcaption}
\usepackage{url}
\usepackage[bookmarks=false]{hyperref}

\usepackage{graphicx}
\usepackage[dvipsnames]{xcolor}
\usepackage{enumitem}

\usepackage{indentfirst}

\usepackage{longtable,tabularx}
\usepackage{multicol}
\usepackage{multirow}
\usepackage{booktabs}

\usepackage{amsfonts}
\usepackage{amsthm}
\usepackage{amssymb}
\usepackage{amsmath}
\usepackage{xfrac}

\usepackage{tikz}
\usetikzlibrary{positioning,decorations.pathreplacing,arrows,arrows.meta,backgrounds,fit,calc,shapes}

\newcommand{\bs}{\boldsymbol}

\title{Predictive Digital Twin for Optimizing Patient-Specific Radiotherapy Regimens under Uncertainty in High-Grade Gliomas}
\author[1]{Anirban Chaudhuri}
\author[1]{Graham Pash}
\author[1,7]{David A. Hormuth II}
\author[1,3]{Guillermo Lorenzo}
\author[1]{Michael Kapteyn}
\author[1]{Chengyue Wu}
\author[1,4]{Ernesto A. B. F. Lima}
\author[1,2,5,6,7,8]{Thomas E. Yankeelov}
\author[1]{Karen Willcox}
\affil[1]{\footnotesize Oden Institute for Computational Engineering and Sciences, The University of Texas at Austin, TX, USA}
\affil[2]{\footnotesize Department of Biomedical Engineering, The University of Texas at Austin, Austin, TX, USA}
\affil[3]{\footnotesize Department of Civil Engineering and Architecture, University of Pavia, Italy}
\affil[4]{\footnotesize Texas Advanced Computing Center, The University of Texas at Austin, Austin, TX, USA}
\affil[5]{\footnotesize Department of Diagnostic Medicine, The University of Texas at Austin, Austin, TX, USA}
\affil[6]{\footnotesize Department of Oncology, The University of Texas at Austin, Austin, TX, USA}
\affil[7]{\footnotesize Livestrong Cancer Institutes, The University of Texas at Austin, Austin, TX, USA}
\affil[8]{\footnotesize Department of Imaging Physics, MD Anderson Cancer Center, Houston, TX, USA}

\date{\vspace{-3em}}

\graphicspath{{figs/}}

\begin{document}
	
	\maketitle
	
	\begin{abstract}
		We develop a methodology to create data-driven predictive digital twins for optimal risk-aware clinical decision-making. We illustrate the methodology as an enabler for an anticipatory personalized treatment that accounts for uncertainties in the underlying tumor biology in high-grade gliomas, where heterogeneity in the response to standard-of-care (SOC) radiotherapy contributes to sub-optimal patient outcomes. The digital twin is initialized through prior distributions derived from population-level clinical data in the literature for a mechanistic model's parameters. Then the digital twin is personalized using Bayesian model calibration for assimilating patient-specific magnetic resonance imaging data. The calibrated digital twin is used to propose optimal radiotherapy treatment regimens by solving a multi-objective risk-based optimization under uncertainty problem. The solution leads to a suite of patient-specific optimal radiotherapy treatment regimens exhibiting varying levels of trade-off between the two competing clinical objectives: (i) maximizing tumor control (characterized by minimizing the risk of tumor volume growth) and (ii) minimizing the toxicity from radiotherapy. The proposed digital twin framework is illustrated by generating an \textit{in silico} cohort of 100 patients with high-grade glioma growth and response properties typically observed in the literature. For the same total radiation dose as the SOC, the personalized treatment regimens lead to median increase in tumor time to progression of around six days. Alternatively, for the same level of tumor control as the SOC, the digital twin provides optimal treatment options that lead to a median reduction in radiation dose by 16.7\% (10 Gy) compared to SOC total dose of 60 Gy. The range of optimal solutions also provide options with increased doses for patients with aggressive cancer, where SOC does not lead to sufficient tumor control.
		
		
	\end{abstract}

	\section{Introduction}\label{s:intro}
\begin{figure}[!htbp]
		\begin{center}
			\includegraphics[width=\textwidth]{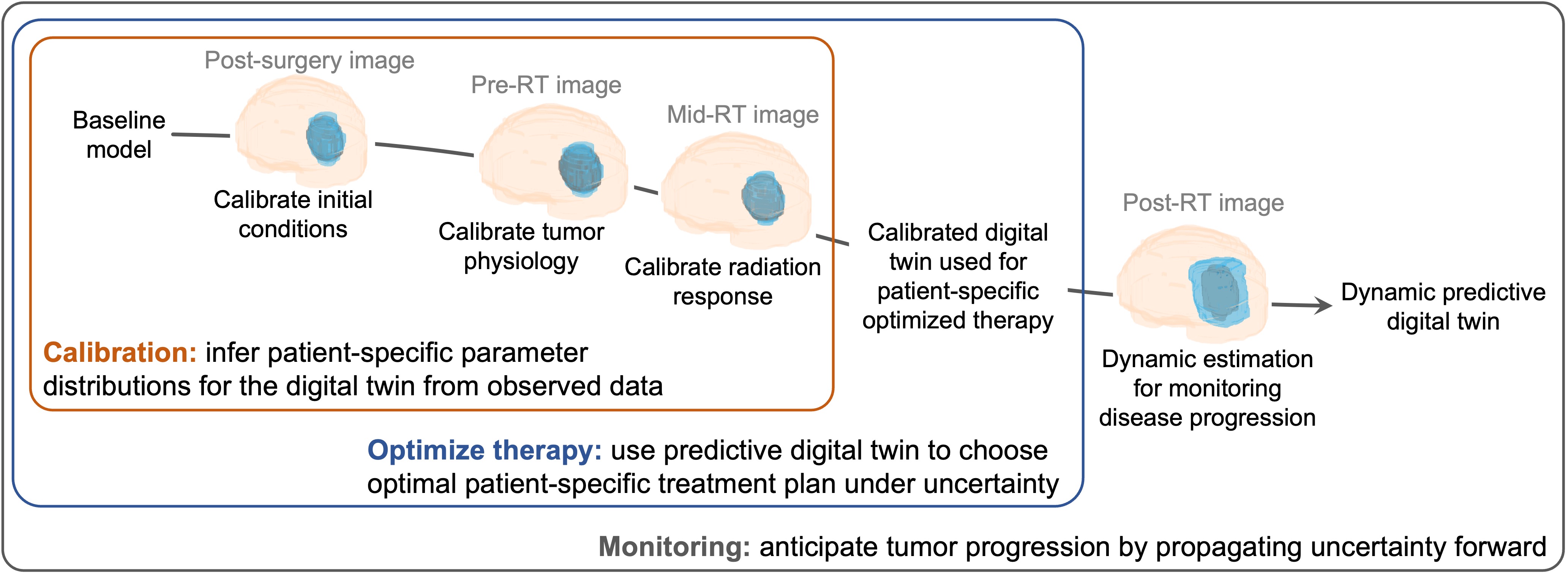}\vspace{0cm}
		\end{center}
		\caption{\textbf{Creating and evolving an HGG patient-specific predictive digital twin.} Our digital twin methodology is illustrated for the case of HGG growth and response to RT from the post-surgery imaging visit to post-RT monitoring for disease progression. The digital twin is personalized through calibration using MRI data and then used to design an optimal risk-aware treatment regimen under uncertainty. The digital twin also allows for monitoring the disease progression throughout the patient's treatment and recovery.}
		\label{fig:dt_onco_summary}
	\end{figure}

	A digital twin can be defined as a mathematical model (or a collection of models) that provides a virtual representation of a specific physical object (e.g., a tumor), updates its status by assimilating object-specific data (e.g., imaging and clinical measurements of tumor growth and radiotherapy response), predicts the behavior of the object under external actions (e.g., treatments), and enables decision-making to optimize the future behavior of the object (e.g., design an optimal radiotherapy plan maximizing tumor control and minimizing toxicities)~\cite{kapteyn2021pgm,Niederer2021,Wu2022DTwin,Rasheed2020,laubenbacher2021using}. Recent work explores the use of digital twins in healthcare and medicine to perform simulations of cardiovascular diseases~\cite{CorralAcero2020,Peirlinck2021}, enable virtual reality for surgery~\cite{Ahmed2020}, and enable improved decision-making in clinical oncology~\cite{madhavan2021envisioning, hernandez2021digital,Wu2022DTwin}. However, most of the work on digital twins in medicine relies on deterministic implementations. Predictive digital twins account for uncertainty through a Bayesian framework~\cite{kapteyn2021pgm} and provide a computational environment to support personalized risk-based management of solid tumors supported by computer forecasts of biologically-inspired mechanistic models representing these diseases and their treatments. We propose a patient-specific predictive digital twin that can address the three critical needs of: (i) accounting for uncertainty in the mechanistic model parameters by continuous integration of incoming patient data, (ii) forecasting the parameter uncertainty to estimate risk associated with the therapeutic outcomes, and (iii) supporting risk-aware clinical decision-making under uncertainty. To illustrate the methodology, we present a predictive digital twin to enable the personalized monitoring and forecast of high-grade glioma (HGG) response to radiotherapy (RT), as well as the design of optimal adaptive RT regimens for individual HGG patients (see Figure~\ref{fig:dt_onco_summary}).
	
	HGG is a class of brain tumors that typically exhibit an aggressive, infiltrative behavior as well as high heterogeneity in both tumor physiology and cell composition~\cite{Omuro2013}. Patients with HGG are usually treated surgical resection to reduce the tumor burden and intracranial pressure followed by adjuvant radiotherapy (RT) and chemotherapy to target residual, unresected tumor cells. The standard-of-care (SOC) RT plans account for patient-specific heterogeneity in tumor shape and location through pre-treatment anatomical or structural imaging approaches, such as $T_1$- and $T_2$-weighted magnetic resonance imaging (MRI), that can identify the residual tumor after surgery and define a surrounding 2--3 cm margin. However, the SOC RT dose and schedule generally conform to the Stupp protocol~\cite{Stupp2005}, which is informed from clinical studies involving large populations, consisting of 60 Gy delivered in 30 fractions of 2 Gy. A fundamental challenge in the treatment of HGG is that response to RT is highly variable from patient to patient due to the inherent heterogeneity in both cellular architecture and tumor micro-environment~\cite{Aum2014,Hill2015}, which may ultimately lead to poor treatment outcomes. This heterogeneity in tumor physiology, growth, and radiation response characteristics translates into uncertainty in the therapeutic outcomes of HGG patients receiving a standard RT protocol. 
	This work addresses the challenge of quantifying the uncertainty in tumor characteristics from limited patient-specific data early in the course of treatment, propagating this uncertainty via mechanistic models to determine the uncertainty in treatment outcomes, and optimizing adaptive RT plans to improve overall survival through individualized predictive digital twins. 
	
	Adaptive RT has been regarded as a promising strategy to account for heterogeneity in tumor response. The varying treatment regimens aim at adjusting the dose, number, and timing of the radiation fractions in the prescribed RT plan on a patient-specific basis to account for the individual response of their HGG to radiation.
	Adaptive RT plans can be guided by quantitative imaging measurements to improve target delineation to deliver radiation, enable the voxelization of the RT plan (i.e., dose-painting), and facilitate the adaptation of the treatment regimen in response to observed tumor dynamics~\cite{Jaffray2012,Troost2015,Kong2017,raaymakers2009integrating}. Another approach for adaptive RT planning is through a predictive framework that can be constructed by leveraging patient-specific computational forecasts of HGG growth and response to RT, which are obtained via computer simulation of biologically-inspired mechanistic models informed by routine clinical and imaging data collected before and during the course of treatment~\cite{Rockne2009,Unkelbach2014,Le2016,Poleszczuk2018,Zahid2021,subramanian2020did,Lorenzo2022,Hormuth21ER,Enderling2019,Woodall2021}.
	However, the tumor forecasts obtained with these models are estimated using a deterministic approach. Systematically accounting for data and model uncertainties has shown promise in Bayesian calibration of tumor models~\cite{oden2017predictive,Lima2017,Lipkova2019,Lorenzo2022}. The clinical translation of current computational technologies to forecast HGG response to RT needs to accommodate a probabilistic risk assessment of pathological and therapeutic outcomes to support clinical decision-making along with progressive update of model according to incoming data. Predictive digital twins are a step towards personalized medicine that can address patient-specific clinical decision-making while accounting for the underlying uncertainty.
	
	We construct the digital twin on an MRI-informed biology-inspired mechanistic model, which describes HGG growth and response to RT, and a probabilistic graphical model~\cite{kapteyn2021pgm}, which accounts for uncertainty quantification and enables the risk assessment of therapeutic outcomes during tumor forecasting and RT optimization. We use prior distributions of the mechanistic model parameters informed from clinical data to initialize the digital twin. The personalization of the digital twin starts with the Bayesian model calibration for assimilating patient-specific MRI data followed by risk-aware optimized RT regimens for each patient. We solve a multi-objective risk-based optimization under uncertainty (OUU) problem to provide a suite of optimal RT regimens that trade-off maximizing tumor control and minimizing the toxicity from RT. We use the $\alpha$-superquantile risk measure~\cite{rockafellar2000optimization} that accounts for the magnitude of exceeding a set threshold in a particular tumor characteristic. The risk-based optimization formulation allows one to account for the patient's and treating physician's risk preferences. We illustrate the predictive digital twin using an \emph{in silico} population of HGG patients that is constructed by pooling clinical data of MRI measurements of HGG cellularity and mechanistic parameter values describing the dynamics of HGG growth and RT response from the literature~\cite{Wang_Rockhill_Mrugala_Peacock_Lai_Jusenius_Wardlaw_Cloughesy_Spence_Rockne_2009,Qi_Schultz_Li_2006,Hormuth21SciRep}. We investigate varying levels of total dose to analyze the ensuing suite of therapeutic planning options. Our results show that the optimal RT plans can lengthen time to progression (TTP) with respect to the SOC regimen for the same total RT dose. Furthermore, we show that for certain patients, the proposed digital twin can provide optimal RT plans achieving similar tumor control as the SOC regimen, while significantly lower toxicities by lowering the radiation dose. We show that the range of optimal solutions also provide options for patients with aggressive cancer with increased radiation doses. We demonstrate that the optimal RT plans are non-inferior to the SOC regimen in terms of tumor control and always maintain the total radiation dose within a clinically-admissible range.
	
	The rest of this work is organized as follows. Section~\ref{s:methods} describes the proposed digital twin methodology to create a patient-specific predictive digital twin for HGG growth and RT response under uncertainty. Section~\ref{s:results} illustrates the digital twin methodology for a cohort of \textit{in silico} patients. Section~\ref{s:discuss} discusses the proposed predictive digital twin framework as well as the results of our computational study, the limitations of this work, and future lines of investigation.

	\section{Patient-specific digital twin methodology}\label{s:methods}
	In this section, we describe the methodologies used to create, update, and utilize a patient-specific predictive digital twin. We begin with an overview of the components comprising the predictive digital twin and ground them in the oncology setting in Section~\ref{s:formulation}. We then describe the mechanistic model in Section~\ref{s:ode} followed by the treatment control parameters in Section~\ref{s:treatment} and the generation of observational data for an \textit{in silico} patient in Section~\ref{s:obs}. The Bayesian model calibration, the propagation of uncertainty for patient-specific prognosis, and the multi-objective risk-based optimization under uncertainty problem are described in Sections~\ref{s:inv}, \ref{s:risk}, and \ref{s:ouu}, respectively. We detail the survival analysis method used to assess the performance of the optimal treatment plans in Section~\ref{s:km}.
	
	\subsection{Predictive digital twin formulation}\label{s:formulation}
\begin{figure}[!htb]
		\begin{center}
			\resizebox{\textwidth}{!}{%
				\input{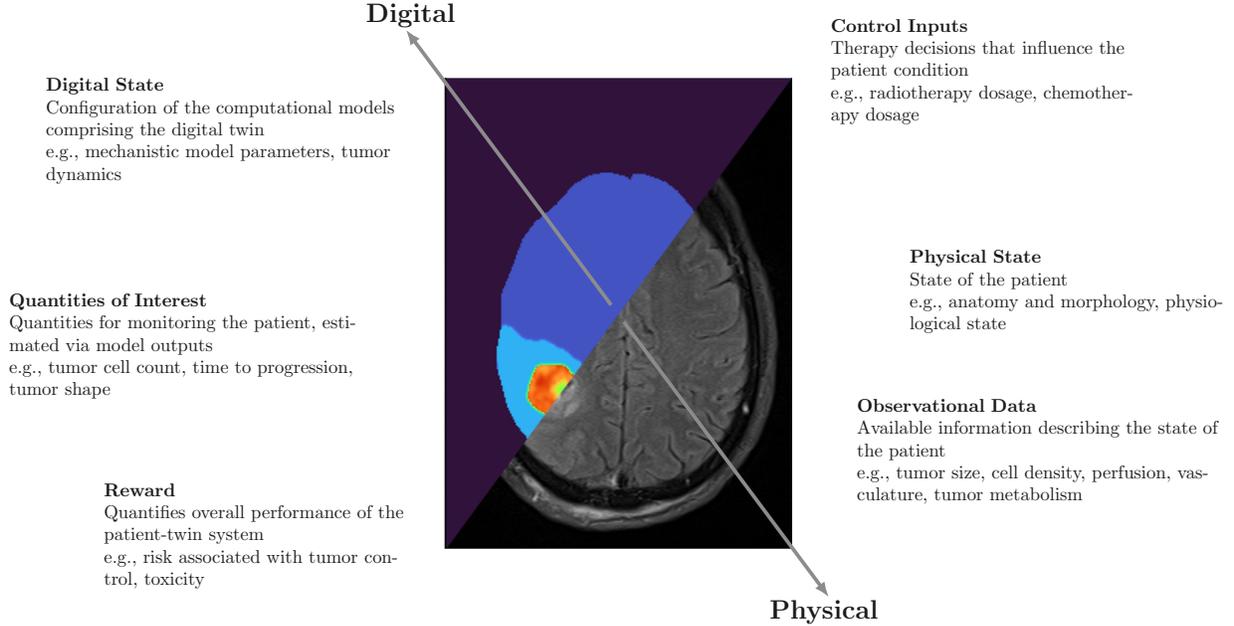}
			}%
		\end{center}
		\caption{Overview of the components comprising a predictive digital twin in the oncology setting.}
		\label{fig:overview}
	\end{figure}

	We adopt the mathematical abstraction for a predictive digital twin proposed in \cite{kapteyn2021pgm}. We formally define a predictive digital twin in terms of the six quantities shown in Figure~\ref{fig:overview} representing the physical and digital twins: physical state, observational data, control inputs, digital state, quantities of interest, and reward, where each of these six quantities will be considered to vary with time. The \textit{physical twin} refers to the specific patient and the \textit{physical state} represents the physiology and anatomy of the patient, which is only partially and indirectly observable. The \textit{digital twin} is characterized by a computational model or a set of coupled computational models that can represent the physical twin to the desired level. The \textit{digital state} is considered to be the parameterization of the computational models comprising the digital twin. To understand the health of the patient and inform our digital state, we rely upon \textit{observational data} obtained from the physical twin, such as data obtained from MRI. There is a strong relationship between the digital state and the observational data since the accuracy of the digital state depends on the type and quantity of the observational data. The \textit{control inputs} in a clinical setting are the therapeutic decisions that influence the physical state of the patient, such as the dose and scheduling of medical interventions. The control inputs can also comprise scheduling decisions for observational data, such as when a patient comes in for imaging. We will use the predictive digital twin to inform our choice of control inputs. The \textit{quantities of interest} (QOIs) are computational estimates of possibly unobservable patient characteristics, which are evaluated using the computational models underlying the digital twin, such as tumor cell count, time to progression and toxicity. The \textit{reward} is used to quantify the performance of the patient-twin system and can encode the success or failure of the system after applying a specific control, such as the risk of under-treating a specific disease giving a therapeutic plan. These quantities describe an abstract coupled patient-twin system representing a patient physical twin and their associated computational digital twin.

	\subsection{Mechanistic tumor growth model comprising the HGG digital twin}\label{s:ode}
	The computational model underlying the predictive digital twin is the logistic growth model of tumor growth governed by the ordinary differential equation (ODE) \cite{Benzekry2014,jarrett_mathematical_2018,hormuth_addr_2022}
	\begin{equation}\label{e:ode}
		\begin{split}
			\frac{\mathrm{d}N}{\mathrm{d}t} &= \rho N\left( 1 - \frac{N}{K} \right);\\ N(t=0)&=N_\text{initial},
		\end{split}
	\end{equation}
	where $N(t)$ is the number of tumor cells at time $t$, $N_\text{initial}$ is the initial tumor burden, $\rho$ is the net proliferation rate of the tumor cells, and $K$ is the carrying capacity of the tissue (i.e.,  the maximum number of tumor cells that can be sustained physically and biologically).
	
	We model the effects of RT and concomitant chemotherapy as discrete treatment events that result in an instantaneous reduction of tumor cell count at treatment time, $N_{\text{post-treatment}}$, to a fraction of the pre-treatment cell count $N_\text{pre-treatment}$, given by
	\begin{equation}
		\label{e:treatment_effect}
		N_{\text{post-treatment}}(t; u_t) = N_\text{pre-treatment}(t)S(u_t),
	\end{equation}
	where $S$ is the surviving fraction of tumor cells resulting from a single dose of RT and chemotherapy at time $t$, which is given by $u_t$. The surviving fraction is defined by a linear-quadratic model of cell survival \cite{McMahon2019} with a multiplicative term to account for the concurrent chemotherapy effect as
	\begin{equation}
		\label{e:survive}
		S(u_t) = S_\text{C} S_\text{RT}(u_t) = S_\text{C} \exp\left(-\alpha_\text{RT} u_t -\beta_\text{RT} u_t^2\right),  
	\end{equation}
	where $S_\text{C}$ is the surviving fraction resulting from a single dose of chemotherapy, $S_\text{RT}$ is the surviving fraction resulting from a single dose of RT,  and $\alpha_\text{RT}$ and $\beta_\text{RT}$ are the radiosensitivity parameters. 
	
	The entire model formulation given by Eqs.~\eqref{e:ode}-\eqref{e:survive} can be written as 
	\begin{equation}
		\frac{\mathrm{d} N}{\mathrm{d} t} = f(N; \theta, \bs{u}),
		\label{e:forward_ode}
	\end{equation}
	where $N(t; \theta,\bs{u})$ is the number of tumor cells at time $t$ given that $\theta:= \left[\rho, K,N_\text{initial},\alpha_\text{RT}\right]^\top \in \Omega \subseteq \mathbb{R}_+^4$ are the patient-specific probabilistic model parameters in the digital state and $\bs{u}$ is a vector comprising all the RT treatment doses $u_t$ given at time $t$. The treatment control considered here adapts the RT dose $\bs{u}$ to specific patients as elaborated in Section~\ref{s:treatment}. All the parameters necessary to define the ODE model are provided in Table~\ref{tab:model_params}. We fix the radiosensitivity parameter ratio to $\sfrac{\alpha_\text{RT}}{\beta_\text{RT}}=10$ \cite{Rockne2009} and the surviving fraction resulting from chemotherapy to $S_\text{C}=0.82$ \cite{Hormuth21SciRep}. The system of ODEs in Eq.~\eqref{e:forward_ode} is solved via a forward Euler scheme with sufficiently small time step size of 0.2 days to ensure numerical stability. The discrete treatment events are applied at the beginning of each day.
	\begin{table}[!htb]
		\centering
		\caption{Mechanistic tumor growth model parameters}
		\label{tab:model_params}
		\begin{tabular}{llcl}
			\toprule
			Parameter type & Name & Symbol & Units \\
			\midrule
			\multirow{4}{*}{Probabilistic parameters} & Proliferation rate & $\rho$ & day$^{-1}$ \\
			& Carrying capacity & $K$ & cells \\
			& Initial tumor burden & $N_\text{initial}$ & cells \\
			& Radiosensitivity parameter & $\alpha_\text{RT}$ & Gy$^{-1}$\\
			\midrule
			\multirow{2}{*}{Fixed parameters} & Surviving fraction due to chemotherapy & $S_\text{C}$  & - \\
			& Radiosensitivity parameter ratio & $\sfrac{\alpha_\text{RT}}{\beta_\text{RT}}$ & Gy \\
			\bottomrule
		\end{tabular}
	\end{table}
	
	\subsection{Treatment control: patient-specific RT treatment regimen}\label{s:treatment}
	The treatment goal considered for this digital twin illustration is to adapt the RT regimen to specific patients. The Stupp protocol \cite{Stupp2005} is the current SOC for treatment of HGG. The SOC administers six weeks of treatment featuring five consecutive treatment days each week with a fractionated RT dose of 2 Gy/day. We consider a similar setup as the SOC dose fractionation to define the treatment control $\bs{u}\in \mathcal{U} \subseteq \mathbb{R}_+^{n_u}$ as a vector of weekly-fractionated RT dose with $n_u=6$ being the number of weeks of RT. Under this setup, the SOC is represented by $\bs{u}_\text{SOC}=[2,2,2,2,2,2]$~Gy/day leading to a total dose of $5\lVert \bs{u}_\text{SOC}\rVert_1=60$ Gy, where $\|\cdot\|_1$ denotes the $\ell_1$ norm (i.e., the summation of the absolute value of the vector entries). The SOC is used as the baseline to compare with the optimized RT treatment regimens. We administer chemotherapy throughout the six weeks even when no RT is prescribed. Note that extensions of this work can consider a finer time scale, such as allowing for a varying dose on a daily basis within each week of the treatment as well as adapting the chemotherapy treatment.
	
	To calibrate the RT-related parameters in the predictive digital twin, we need at least one observation after starting the RT treatment (see Section~\ref{s:inv} for details on calibration). Thus, we consider that the patients receive the SOC RT plan during this first week and that the intra-treatment MRI is performed at the end of the first week of RT (here, simulated patients as described in Section~\ref{s:obs}). We fix the fractionated dose of the first week to the SOC value of 2~Gy/day and use the predictive digital twin to optimally control the RT doses for the remaining five weeks (i.e., $\bs{u}=[u_1=2,u_2,\dots,u_6]$, where $u_i \in [0,10]$~Gy/day, $i=2,\dots,6$). Importantly, this strategy to optimally adjust the RT plan with our predictive digital twin follows the same approach as image-driven adaptive RT methods that are currently being explored in the clinical-research setting \cite{Nijkamp,Sonke}.  
	
	\subsection{Simulated patients as proxy for observational clinical data}\label{s:obs}
	The physical twin state of each patient is partially and indirectly observed through specific observational data, which are leveraged to inform the digital twin state. In the HGG setting, these observational data can be acquired non-invasively using MRI. Recently, MRI data have been used to calibrate computational models and obtain tumor forecasts~\cite{hormuth_pmb,Hormuth20RadOnc,Hormuth21Cancers,Hormuth21SciRep} after appropriate post-processing. In particular, the MRI data needs to be post-processed to extract the total tumor burden as a cell count, such that it relates to the state variable $N$ of the ODE model in Eq.~\eqref{e:forward_ode}. Since we are collapsing the spatial information of the tumor provided by the MRI data, we incur a volume-wise error in the measurement of the tumor burden. This source of error has been studied by both \cite{paldino2014repeatability} and \cite{mazzara2004brain}, although it is difficult to quantify its effect on the model parameterization and ensuing forecasts.
	
	We consider an {\it in silico} patient cohort where the physical state of the HGG tumors is simulated by generating noisy observations of the solution to the ODE model in Eq.~\eqref{e:forward_ode} using an underlying ``true" parameter set $\theta_\text{true}$, which is varied for each patient in the cohort. Note that $\theta_\text{true}$ is never seen by the predictive digital twin and the model parameters instead adopt a probabilistic formulation based on the noisy observations generated for each \textit{in silico} patient. We simulate a collection of $n_\text{obs}$ observations $\bs{o}=\left[o_{t_1},\dots,o_{t_{n_\text{obs}}}\right]$ at times $\{t_i\}_{i=1}^{n_\text{obs}}$ with an additive noise model as
	\begin{equation}\label{e:noise}
		o_{t_i} = N(t_i; \theta_\text{true},\bs{u}) + \varepsilon_i,
	\end{equation}
	where the noise $\varepsilon_i$ follows a truncated normal distribution $\mathcal{TN}(0, \sigma^2, -N(t_i; \theta_\text{true},\bs{u}), +\infty)$ with a lower truncation bound of $-N(t_i; \theta_\text{true},\bs{u})$ to avoid non-physical negative observations. The general truncated normal distribution definition $\mathcal{TN}(\mu, \sigma^2, a, b)$ can be read as $\mu$ and $\sigma^2$ specifying the mean and variance, respectively, of the general normal distribution with the truncation range as $[a,b]$~\cite{cohen1991truncated}. For this work, we assume a constant standard deviation of $\sigma = 2\times 10^9$ cells, corresponding to a value of 10\% of the mean value from population data used to model the initial tumor burden.

	\subsection{Patient-specific tumor modeling via Bayesian model calibration} \label{s:inv}
	We use a Bayesian formulation that combines prior knowledge with observed data to update the probabilistic distribution placed on the model parameters. The observational data from the patient are assimilated by solving an inverse problem to calibrate the parameters in the computational model \cite{biros2011large, stuart2010inverse,oden2016toward} to the specific patient. The Bayesian framework provides confidence levels for the computational model output. Prior knowledge plays a critical role in the oncology setting as it is an especially data-poor regime owing to the difficulty and expense of collecting quantitative patient information such as MRI data. Priors allow us to inject knowledge of the disease at the population level to inform the modeling process of a specific patient. Priors $\mathcal{P}(\theta)$ for the probabilistic parameters $\theta$ are constructed from reported values of clinical data in the literature. The study by~\cite{Wang_Rockhill_Mrugala_Peacock_Lai_Jusenius_Wardlaw_Cloughesy_Spence_Rockne_2009} computed the global proliferation rate for a cohort of 31 patients. \cite{Qi_Schultz_Li_2006} conducted a study to compute a plausible set of radio-sensitivity parameters for the linear-quadratic model.  We define an independent truncated normal distribution based on the values from the literature as our prior distribution as shown in Table~\ref{tab:prior}. A truncated normal distribution is used to enforce positivity as well as to account for physically reasonable upper and lower bounds on the model parameters. 
	\begin{table}[!htb]
		\centering
		\caption{Prior distribution for the probabilistic model parameters $\theta:= \left[\rho, K,N_\text{initial},\alpha_\text{RT}\right]^\top$ }
		\label{tab:prior}
		\begin{tabular}{ccccc}
			\toprule
			Parameter & Mean & Standard deviation & Lower bound & Upper bound \\
			\midrule
			$\rho$ & 0.09 & 0.15 & 0.007 & 0.25 \\
			$K$ & $1\times 10^{11}$ & $2\times 10^{10}$ & $9\times 10^{10}$ & $1.8\times 10^{11}$  \\
			$N_\text{initial}$ & $1.9\times 10^{10}$ & $1.2\times 10^{10}$ & $4.7\times 10^{9}$ & $4.7\times 10^{10}$ \\
			$\alpha_\text{RT}$ & 0.05 & 0.025 & 0.001 & 0.1 \\
			\bottomrule
		\end{tabular}
	\end{table}
	
	Observational data $\bs{o}$ from the patient are assimilated to estimate the updated probability distribution of $\theta$, otherwise known as the posterior distribution, through the Bayesian update formula given by
	\begin{equation}\label{e:bayes}
		\mathcal{P}(\theta | \bs{o}) \propto \mathcal{P}(\bs{o} | \theta) \mathcal{P}(\theta),
	\end{equation}
	where $\mathcal{P}(\theta | \bs{o})$ is the posterior distribution of the digital state conditioned upon the observed data, $\mathcal{P}(\bs{o} | \theta)$ is the likelihood of the observed data given a digital state, and $\mathcal{P}(\theta)$ is the prior distribution that encodes our knowledge about the patient's state before any observations are made.
	
	We consider a scenario with three available MRI observations on days 0 (post-surgery image), 20 (pre-RT image), and 27 (mid-RT image). As noted in Eq. \eqref{e:noise}, we assume an additive noise model for our observational data generation. The observation on day 0 only provides knowledge about the initial tumor burden $N_\text{initial}$ for the patient. The observation on day 20 provides information on all the probabilistic model parameters except the radiosensitivity parameter $\alpha_\text{RT}$. The observation on day 27 is generated using the SOC fractionated dose and provides knowledge about all the probabilistic model parameters $\{\rho, K,N_\text{initial},\alpha_\text{RT}\}$. This leads us to solve the inverse problem for model calibration in two steps:
	\begin{description}[itemsep=1pt,topsep=0pt]
		\item[\textbf{Step 1: }] After obtaining the post-surgery image for the patient, the first step updates the posterior of the initial tumor burden to use the truncated normal distribution associated with the observation $o_{t_1}$ at $t_1=0$ days as the mean with $\sigma^2$ accounting for the measurement error as $N_\text{initial} \sim \mathcal{TN}(o_{t_1}, \sigma^2, \max\{0,o_{t_1}-2\sigma\}, o_{t_1}+2\sigma)$. The upper and lower truncation bounds are defined by $\pm2\sigma$ from the observational data. The lower truncation bound is further modified to avoid non-physical negative values of initial tumor burden by using $\max\{0,o_{t_1}-2\sigma\}$. The updated distribution of $N_\text{initial}$ is used as the prior distribution in Step 2.
		\item[\textbf{Step 2: }] The Bayesian inference is performed via Markov Chain Monte Carlo (MCMC) for assimilating the remaining two observations $[o_{t_2}, o_{t_3}]$ from $t_2=20$ days and $t_3=27$ days to complete the digital twin model calibration for tumor physiology and radiation response. We use the updated distribution of $N_\text{initial}$ from step 1 and the priors for $\rho, K,\alpha_\text{RT}$ given in Table~\ref{tab:prior}. Then the patient-specific posterior distribution for all the probabilistic model parameters are obtained by using MCMC for the remaining two observations. We use the MCMC implementation in the PyMC3 Python package~\cite{Salvatier2016} with four chains of $100,000$ samples each. The posterior distribution is characterized by $100,000$ retained samples. For a more in-depth discussion of MCMC techniques, the reader is referred to~\cite{smith2013uncertainty} and the references within.
	\end{description}
	
	\subsection{Patient-specific prognosis via  uncertainty propagation}\label{s:risk}
	Consider an uncertain digital state defined by the posterior distributions of the model parameters generated by Bayesian model calibration as described in Section~\ref{s:inv}. At any point in time, we can propagate uncertainty forward in time using the model in Eq.~\eqref{e:forward_ode} to estimate our quantities of interest, which characterize the tumor control resulting from an RT regimen and the treatment toxicities. Hence, a QoI maps realizations of random model parameters $\theta\in\Omega$ and a treatment control $\bs{u}\in\mathcal{U}$ to a real value as $M:\mathcal{U}\times\Omega \mapsto \mathbb{R}$. There are a variety of clinically relevant QOIs that could be considered to characterize the tumor control resulting from a treatment regimen. These include the predicted tumor burden (e.g., defined through the tumor cell count) after a set period of time post-treatment, time to progression (TTP) of the tumor cell count beyond a given threshold, or the amount of time that the tumor cell count is kept below a specified threshold. In this work, the QoI for tumor control is based on the TTP~\cite{Saad_Katz_2009}. We consider toxicity to be proportional to the total dose of radiation delivered to the patient during RT, i.e, $5 \lVert \bs{u} \rVert_1$. Note that the QoI for toxicity does not depend on $\theta$ and, thus, it is a deterministic quantity since we have no uncertainty in the dose administered to a patient.
	
	We define the TTP as the amount of time that the tumor takes to proliferate to a clinically-relevant size after the conclusion of six weeks of RT, which we set as the tumor cell count right before the onset of the RT regimen (i.e., at $t=20$ days). Hence, to calculate the TTP, we further define a threshold tumor cell count $N_\text{th}(\theta):=N(t=20;\theta, \bs{u})$, which does not depend on $\bs{u}$ since it is a pre-treatment quantity. Additionally, we denote the day of the conclusion of RT as $t_\text{post-RT}$ (here, $t_\text{post-RT}=20+6\times 7=62$ days). Since a second round of chemotherapy is generally prescribed to prevent or combat tumor progression three months after the conclusion of RT, we consider a finite time-horizon of $t_\text{finite}=152$ days after surgery for the end of our simulations and the calculation of the TTP. We can now express the TTP  as
	\begin{equation}
		\label{e:ttp}
		T_\text{TTP}(\bs{u}, \theta) := \min_{t_\text{post-RT}<t\le t_\text{finite}} \{t :\, N(t; \theta, {\bs u}) > N_\text{th}(\theta) \}-20.
	\end{equation}
	We want to maximize the TTP, or equivalently, minimize the negative of TTP. Thus, we define our QoI for tumor control as $M(\bs{u}, \theta)=-T_\text{TTP}(\bs{u}, \theta)$ to be compatible with the definition of a general minimization problem (see Section~\ref{s:ouu}).
	
\begin{figure}[h!]
		\begin{subfigure}{0.98\textwidth}
			\centering
			\includegraphics[width=\textwidth, page=1]{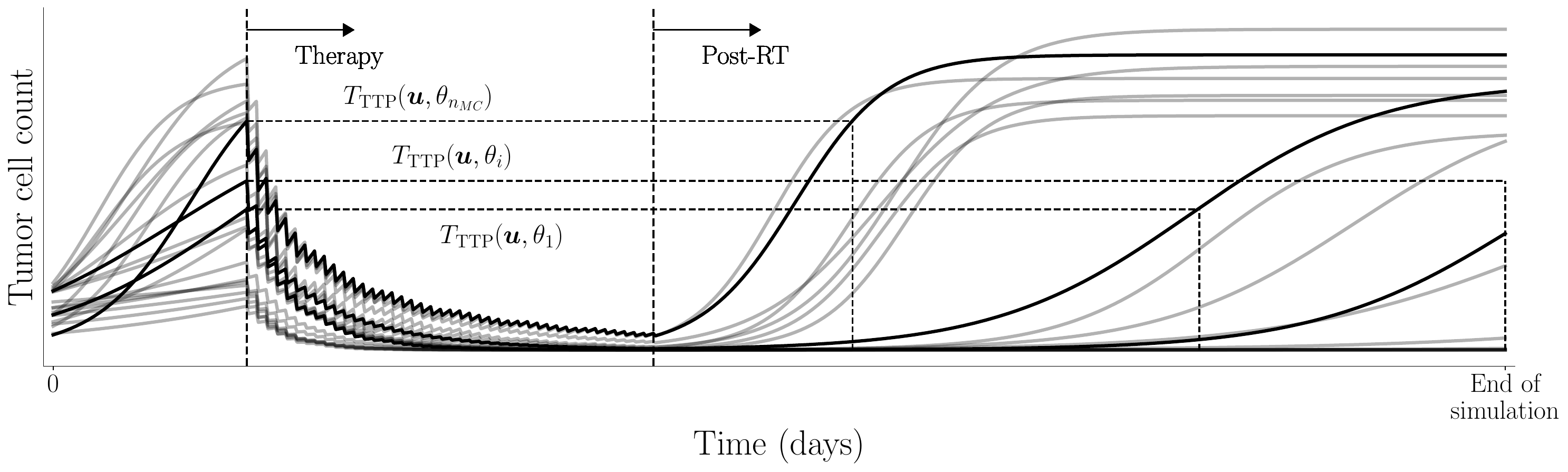}
			\caption{}
			\label{fig:ttp_viz}
		\end{subfigure} \hfill
		\begin{subfigure}{0.49\textwidth}
			\centering
			\includegraphics[width=\textwidth]{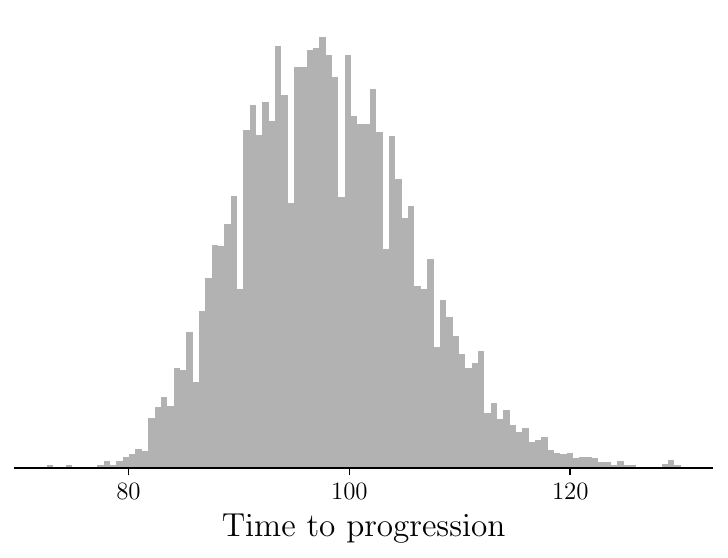}
			\caption{}
			\label{fig:ttp_hist}
		\end{subfigure} \hfill
		\begin{subfigure}{0.49\textwidth}
			\centering
			\includegraphics[width=\textwidth]{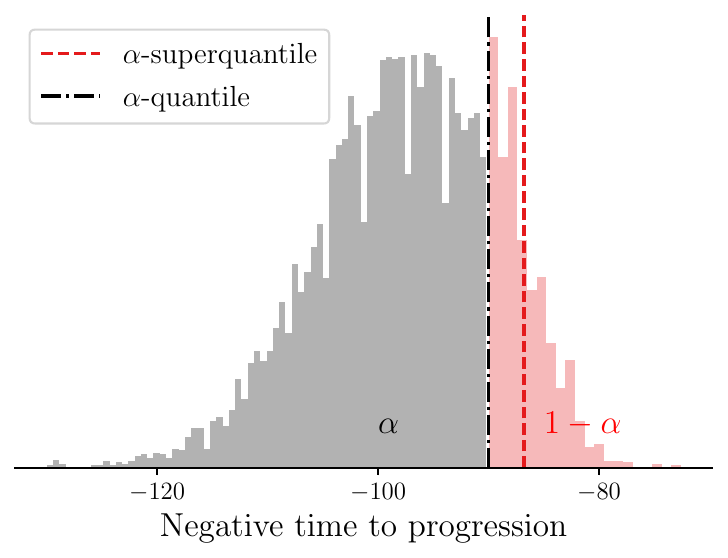}
			\caption{}
			\label{fig:qoi}
		\end{subfigure}
		\caption{\textbf{Illustration of TTP and use of negative TTP as QoI.} (a) Visual representation of TTP for various trajectories obtained from $n_\text{MC}$ Monte Carlo samples of $\theta$, (b) histogram of samples of $T_\text{TTP}(\bs{u}, \theta)$ and (b) histogram of samples of QoI $-T_\text{TTP}(\bs{u}, \theta)$ along with the estimated risk with $\alpha=0.95$.}
		\label{fig:ttp}
	\end{figure}

	In general, the nonlinear effects of the underlying computational model and the non-parametric distributions of $\theta$ prohibit one from obtaining an analytic expression for the distribution of the QoI, hence we rely on the Monte Carlo sampling approach. We sample a realization of the parameter set $\theta$ from the posterior distribution describing a patient's digital state, solve the forward model in Eq.~\eqref{e:forward_ode} to obtain the TTP, and compute the QoI $M(\bs{u}, \theta)$. The process is continued till a desired number of samples $n_\text{MC}$ of $\theta$ are processed to obtain the realizations of TTP as shown in Figure~\ref{fig:ttp_viz}. The distribution of TTP and the connection with $M(\bs{u}, \theta)$ is illustrated in Figures~\ref{fig:ttp_hist} and \ref{fig:qoi}.
	
	To make decisions concerning the patient-specific design of optimal treatment regimens, there are different statistics associated with the distribution of $M(\bs{u}, \theta)$ that can be used as the reward function. We use a risk measure as the statistic to characterize the risk in tumor control associated with a given treatment regimen $\bs{u}$. The risk measure is denoted by $\mathcal{R}:\mathcal{U}\times\Omega \mapsto \mathbb{R}$. Different types of risk measures can be used to quantify the risk associated with tumor control, such as the probability of exceeding a threshold TTP value and the $\alpha$-quantile for a set risk level $\alpha$. We use the $\alpha$-superquantile risk measure~\cite{rockafellar2000optimization,Rockafellar_Royset_2013,rockafellar2015engineering,kouri2016risk}, which has certain desirable mathematical properties. The $\alpha$-superquantile satisfies two notions of certifiability in risk-based OUU~\cite{Chaudhuri_Kramer_Norton_Royset_Willcox_2022}: (i) accounting for near-failure and catastrophic failure events, and (ii) preserving convexity of the underlying QoI to aid in convergence guarantees for risk-based optimization formulations. The $\alpha$-superquantile risk measure accounts for the magnitude of the largest $100(1-\alpha)\%$ realizations that captures the specified portion of the worst-case scenarios through risk level $\alpha$. In oncology, it is important to account for the extent to which a tumor progresses beyond a specified threshold or remains in remission but close to the threshold rather than just the frequency of exceeding the threshold. Clinical ramifications of such extreme cases could result in under-treatment of aggressive disease resulting in poor tumor control and early disease progression. This is especially important in high-grade glioma where the prognosis is already overwhelming poor. The $\alpha$-superquantile risk measure can take into account the magnitude of the TTP exceeding a set threshold to build in a desired level of conservativeness in a data-driven way.
	
	To define the $\alpha$-superquantile risk measure, we first define the related quantity of $\alpha$-quantile $Q_\alpha$ for risk level $\alpha\in(0,1)$ as
	\begin{equation}\label{e:pof}
		Q_\alpha \left[M(\bs{u}, \theta) \right] := F^{-1}_{M(\bs{u}, \theta)}(\alpha),
	\end{equation}
	where $F^{-1}_{M(\bs{u}, \theta)}$ is the inverse cumulative distribution function of $M(\bs{u}, \theta)$. Then, we can define the $\alpha$-superquantile $\overline{Q}_{\alpha}[M(\bs{u}, \theta)]$ as
	\begin{equation}\label{e:cvar}
		\overline{Q}_{\alpha}[M(\bs{u}, \theta)] := Q_\alpha [M(\bs{u}, \theta)] + \frac{1}{1-\alpha} \mathbb{E}\left[\left[M(\bs{u}, \theta) - Q_\alpha \left[M(\bs{u}, \theta) \right]\right]^+\right],
	\end{equation}
	where $[\cdot ]^+ = \max(0, \cdot)$. When the cumulative distribution of $M(\bs{u}, \theta)$ is continuous, we can view $\overline{Q}_{\alpha}[M(\bs{u}, \theta)]$ as the conditional expectation of the QoI conditioned on the QoI exceeding the $\alpha$-quantile as $\overline{Q}_{\alpha}[M(\bs{u}, \theta)] = \mathbb{E}\left[ M(\bs{u}, \theta) \mid M(\bs{u}, \theta) \ge Q_\alpha \left[M(\bs{u}, \theta)\right] \right]$. We use Algorithm 1 in Ref.~\cite{Chaudhuri_Kramer_Norton_Royset_Willcox_2022} to estimate the $\alpha$-superquantile through Monte Carlo simulations. Figure~\ref{fig:qoi} shows an illustration of the data-driven conservativeness inferred from the Monte Carlo samples when using $\alpha$-superquantile compared to $\alpha$-quantile ($\overline{Q}_{\alpha}[M(\bs{u}, \theta)]>Q_{\alpha}[M(\bs{u}, \theta)]$).
	
	The $\alpha$-superquantile risk measure is used to formulate a risk-based optimization problem under uncertainty  to select the optimal treatment regimen as described in Section~\ref{s:ouu}. The $\alpha$-superquantile risk measure helps in making risk-aware decisions, where the risk preference for each patient can be adjusted by changing the value of $\alpha$. Hence, larger values of $\alpha$ lead to more risk-averse decisions. We demonstrate our approach with a value of $\alpha=0.95$.
	
	\subsection{Optimal patient-specific RT treatment regimens under uncertainty}\label{s:ouu}
\begin{figure}[!htb]
		\centering
		\includegraphics[width=\textwidth]{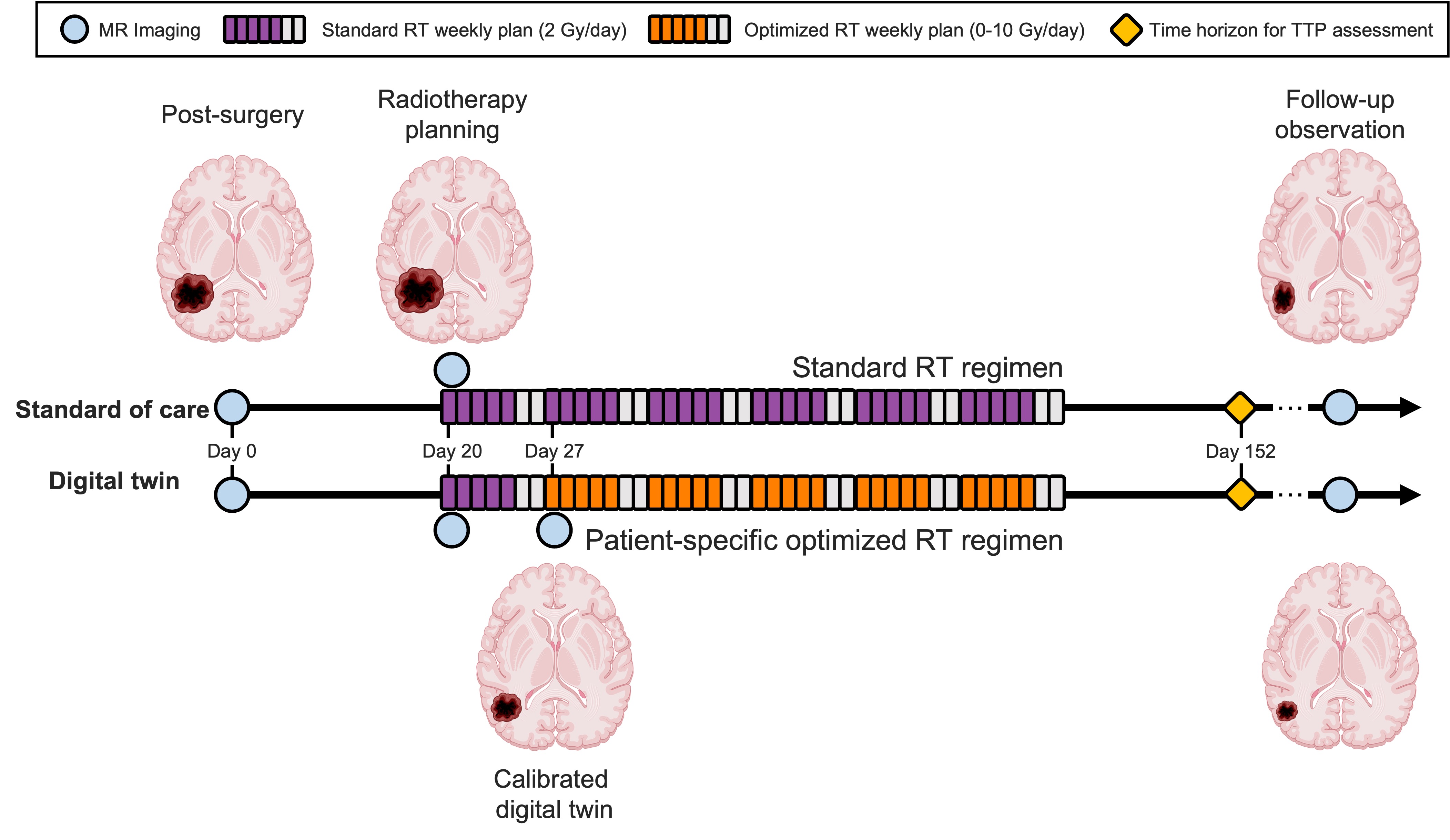}
		\caption{\textbf{Predictive digital twin timeline for HGG patients.} The predictive digital twin features a personalized risk-based strategy to optimize the adaptive RT regimen under uncertainty. This strategy is constructed assuming a standard collection of MRI data during the course of the clinical management of HGG after surgery, whereby MRI scans are prescribed after the surgical intervention (day 0), before the onset of RT (day 20), and during the second week of the RT regimen (day 27). On day 27, the digital twin is calibrated using the MRI data and then the risk-aware treatment plan is solved using the calibrated model and deployed for the remaining five weeks of RT.}
		\label{fig:timeline}
	\end{figure}

	In this work, we consider two clinical objectives: (i) minimizing the risk associated with the QoI characterizing the tumor control, and (ii) minimizing toxicity through a proportional quantity to the total RT dose. To define personalized optimal RT regimens achieving these goals, we formulate a multi-objective risk-based optimization problem under uncertainty. This method results in a suite of eligible optimal RT regimens featuring different levels of trade-off between (i) decreasing toxicity by minimizing the total dose, or (ii) minimizing the risk associated with tumor control. The multi-objective formulation is given by
	\begin{equation}
		\label{e:multiobj_formulation1}
		\bs{u}^* = \underset{\bs{u}\in\mathcal{U}}{\arg\min}\ \left(\mathcal{R}( M(\bs{u}, \theta)), 5\lVert \bs{u}\rVert_1\right),
	\end{equation}
	where optimized therapy comes into effect after the first week of RT and we optimize the RT doses for the remaining five weeks as denoted by $\bs{u}=[u_1=2,u_2,\dots,u_6]$ (see  Section~\ref{s:treatment}). Note that we can also explore the trade-off with changing risk preferences by varying the risk level $\alpha$ for the $\alpha$-superquantile risk measure, as discussed in Section~\ref{s:risk}. An illustration of the predictive digital twin for HGG that shows the specific timeline is shown in Figure~\ref{fig:timeline}.
	
	To solve the multi-objective problem in Eq.~\eqref{e:multiobj_formulation1}, we reformulate it into a constrained single-objective problem using the $\epsilon$-constraint method~\cite{haimes1971bicriterion,hwang2012multiple} as 
	\begin{equation}
		\label{e:multiobj_formulation2}
		\begin{split}
			\bs{u}^* = \underset{\bs{u}\in\mathcal{U}}{\arg\min}\ & \mathcal{R}( M(\bs{u}, \theta))   \\
			\text{s.t. } & 5\lVert \bs{u}\rVert_1 \le D_\text{max},
		\end{split}
	\end{equation}
	where $D_\text{max}$ is the total dose to be delivered. The value of $D_\text{max}$ is varied over a range of total doses to obtain the Pareto optimal solutions~\cite{hwang2012multiple}. Each of the Pareto optimal solutions provide different balance of toxicity and tumor control. Hence, the $\epsilon$-constraint method is suitable for our application since we have a known primary objective of minimizing the risk associated with tumor control, while reducing the dose is a secondary objective for controlling toxicity. We also have a structured way to define the range of values for $D_\text{max}$ based on the preferred total dose, which solves the challenge of selecting meaningful levels in the $\epsilon$-constraint method. Thus, the specific advantages of the $\epsilon$-constraint method in the context of the multi-objective problem defined in Eq.~\eqref{e:multiobj_formulation1} are: (i) preserving properties of the underlying problem, such as convexity of QOIs and risk measures, and (ii) ease of specification of the number of Pareto optimal treatment plans by selecting the number of $D_\text{max}$ values and solving a single-objective constrained optimization problem each time.
	
	The solution of Eq.~\eqref{e:multiobj_formulation2} will always lead to an optimal treatment regimen with active constraint ($5\lVert \bs{u}^*\rVert_1 = D_\text{max}$) since increasing dose directly leads to decreasing risk. However, after an initial analysis of the optimization problem in Eq.~\eqref{e:multiobj_formulation2}, we found that it leads to sub-optimal solutions for patients whose HGG exhibits a low tumor cell proliferation rate and/or low initial tumor burden. We observed that these patients do not require the maximum total dose enforced by an active constraint at the optimal solution to achieve the same amount of tumor control after the conclusion of the RT regimen. Typically for such patients, the amount of tumor control corresponds to a maximum possible value of 132 days for the TTP realizations based on the finite time-horizon of 152 days for our simulations. In other words, the same amount of risk associated with tumor control can be obtained while reducing the total dose below $D_\text{max}$ for these patients. To account for such cases, we add an additional penalty term using parameter $\lambda$ (here, $\lambda=0.001$) on the total delivered dose as
	\begin{equation}
		\label{e:multiobj_formulation}
		\begin{split}
			\bs{u}^* = \underset{\bs{u}\in\mathcal{U}}{\arg\min}\ & \mathcal{R}( M(\bs{u}, \theta))+ \lambda \lVert \bs{u}\rVert_1   \\
			\text{s.t. } & 5\lVert \bs{u}\rVert_1 \le D_\text{max}.
		\end{split}
	\end{equation}
	The single-objective constrained optimization problem in Eq.~\eqref{e:multiobj_formulation} is solved for various selections of the total dose threshold, $D_\text{max}\in \{40, 50, 60, 70, 80, 100\}$ Gy, to obtain the Pareto front of optimal solutions. The SciPy Python package~\cite{virtanen2020scipy} is used for solving the optimization problem. We use the basin-hopping algorithm~\cite{wales1997global} for multi-start local optimization with the gradient-free COBYLA (constrained optimization by linear approximation) optimizer~\cite{powell1994direct} from the SciPy package~\footnote{\href{https://docs.scipy.org/doc/scipy/reference/generated/scipy.optimize.basinhopping.html}{SciPy Basin-hopping}; \href{https://docs.scipy.org/doc/scipy/reference/generated/scipy.optimize.fmin_cobyla.html}{SciPy COBYLA}}. We use 20 restarts of COBYLA with a maximum of 200 function evaluations in each restart. The risk estimation in each function evaluation during the optimization uses $n_\text{MC}=5000$ forward simulations.
	
	The risk-based multi-objective problem formulation with the specific choice of objectives used here has the following desirable properties:
	\begin{itemize}[nosep, itemsep=0pt, labelsep=0.5em, leftmargin=*]
		\item accounts for uncertainty during clinical decision-making process to provide multiple patient-specific optimal treatment regimens
		\item leads to a suite of patient-specific optimal treatment regimens with different levels of trade-offs between tumor control and toxicity from RT to allow more flexibility to consider the patient's and the treating physician's preferences
		\item provides an optimal treatment regimen option for a patient that can achieve similar tumor control as the SOC, but with reduced RT dose to mitigate toxicity effects whenever possible
		\item provides an optimal treatment regimen option with increased RT dose for patients with aggressive cancer, where the SOC does not provide sufficient tumor control either for patient preference or to allow for further surgical intervention
	\end{itemize}
	
	\subsection{Survival analysis}\label{s:km}
	To assess the tumor control achieved with the different optimized treatment regimens for varying $D_\text{max}$ and compare them against the SOC, we analyze the time to tumor progression across the \textit{in silico} patient cohort using the Kaplan-Meier estimator~\cite{kaplan1958nonparametric,d2021methods}. In this case, a right-censor criterion is enforced to account for the finite time-horizon of HGG response simulation at $t_\text{finite}=152$ days, which leads to a maximum TTP value of 132 days. For a cohort of $n_\text{p}$ patients, we define the probability of survival to tumor progression at time $t$ as the probability of the $\alpha$-superquantile value of the TTP being longer than time $t$. Thus, for each treatment obtained with a different $D_\text{max}$, the $\alpha$-superquantile value of the TTP of each patient is used to calculate the survival probability as
	\begin{equation}
		\label{e:km}
		P_\text{S}(t) := \prod_{i:t_i\le t_\text{finite}} \left( 1-\frac{d(t_i)}{m(t_i)}\right),
	\end{equation}
	where $d(t_i)$ denotes the number of patients with TTP equal to time $t_i$ and $m(t_i)$ are the number of patients with TTP larger than time $t_{i-1}$ for all $i=0,\dots,n_\text{p}$. A further advantage of the probabilistic modeling approach used in the predictive digital twin is the ability to estimate the variance in survival probability numerically from the samples of the TTP instead of relying on the approximation provided by the Greenwood's formula~\cite{greenwood1926report}. 
	
	To assess the statistical significance of different treatment plans, the logrank test~\cite{bland2004logrank} is used to compare the survival distributions between digital-twin-based treatment plans and the SOC. The logrank test is implemented through the LifeLines Python package~\cite{davidson2019lifelines}. Note that the null hypothesis for this test is that there is no difference between the populations in the probability of survival, so significant p-values indicate that the considered treatment is significantly different from the SOC.

	\section{Illustrative cohort of \textit{in silico} patients} \label{s:results}
	This section presents an illustration of the digital twin methodology for a cohort of \textit{in silico} patients. We describe the details of the computational study in Section~\ref{s:study} followed by the results for Bayesian calibration (Section~\ref{s:res_cal}), risk-based OUU (Section~\ref{s:res_ouu}), and survival analysis (Section~\ref{s:res_survival}). The code used for generating the results reported for the predictive digital twin is available at \url{https://github.com/Willcox-Research-Group/predictive-dtwin-glioma-frontiers}.
	
	\subsection{Computational study format} \label{s:study}
	We initialize an \textit{in silico} cohort of 100 patients whose true physical states $\theta_\text{true}$ are determined by sampling from the marginal distributions of the population level priors (see Section~\ref{s:obs}). Noisy measurements are then generated from this cohort using Eq.~\eqref{e:noise} to mimic the data that would be collected in the clinic. The specific observational and control timeline used for our predictive digital twins is outlined in Figure~\ref{fig:timeline}. Note that we are considering a finite time-horizon of three months post-RT ($t_\text{finite}=152$ days) for our simulations and optimization. Thus, the maximum possible value of TTP is 132 days (see Eq.~\eqref{e:ttp}).
	
	We divide the cohort of 100 patients into three groups based on the SOC TTP $\alpha$-superquantile exhibited by the patient:
	\begin{enumerate}[nosep, label=(\roman*), widest=(iii), labelsep=0.5em, leftmargin=*]
		\item Early progressors: 16 patients with SOC TTP $\alpha$-superquantile $\le 1$ month after end of RT
		\item Intermediate progressors: 62 patients with SOC TTP $\alpha$-superquantile between $1$-$3$ months after end of RT
		\item Late progressors: 22 patients with SOC TTP $\alpha$-superquantile $\ge 3$ months after end of RT
	\end{enumerate}
	We first analyze the results for Bayesian calibration and risk-aware treatment plans produced by the patient-specific predictive digital twins for three patients with varied growth and response behaviors. The underlying physical state of the three patients is provided in Table~\ref{tab:case_params}. Note that these patient parameters are not known to the digital twin and are only used to generate observational data for the case-study. Patients 1 and 2 are intermediate progressors and Patient 3 is an early progressor according to the patient groups defined above. We then analyze the effectiveness of our predictive digital twins over the 100 patient cohort and the three patient groups.
	\begin{table}[!htb]
		\centering
		\caption{Physical state model parameters $\theta_\text{true}$ for simulating \textit{in silico} case-study patients.}
		\label{tab:case_params}
		\begin{tabular}{lccc}
			\toprule
			Model parameter & Patient 1 & Patient 2 & Patient 3 \\
			\midrule
			Proliferation rate $\rho$ (day$^{-1}$) & 1.14e-01 & 1.09e-01 & 2.25e-01\\
			Carrying capacity $K$ (cells) & 1.17e+11 & 1.09e+11 & 1.40e+11 \\
			Initial tumor burden $N_\text{initial}$ (cells) & 1.54e+10 & 2.60e+10 & 2.62e+10\\
			Radiosensitivity parameter $\alpha$ (Gy$^{-1}$) & 1.05e-03 & 4.58e-02 & 3.90e-02\\
			\midrule
			Chemotherapy surviving fraction $S_\text{C}$ & \multicolumn{3}{c}{0.82}\\
			Radiosensitivity parameter ratio $\alpha/\beta$ (Gy) & \multicolumn{3}{c}{10} \\ \bottomrule
		\end{tabular}
	\end{table}

	\subsection{Patient-specific Bayesian calibration of the digital twin}\label{s:res_cal}
\begin{figure}[!htb]
		\centering
		\begin{subfigure}{0.9\textwidth}
			\includegraphics[width=\linewidth,page=1]{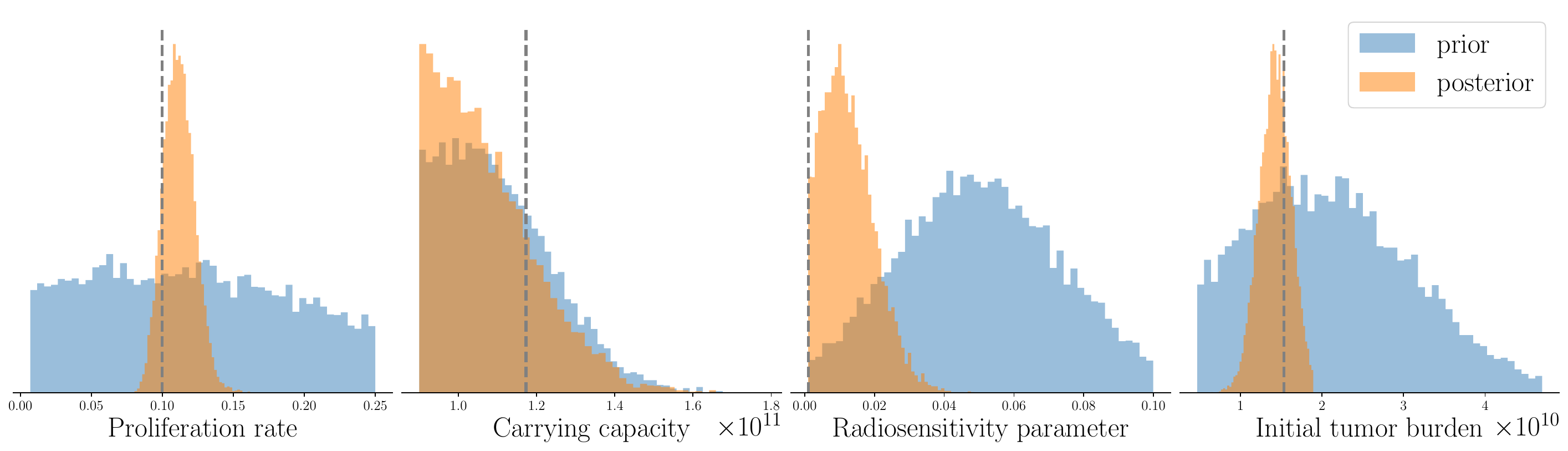}
			\caption{Patient 1.}
			\label{fig:p1inv_dist}
		\end{subfigure}
		\begin{subfigure}{0.9\textwidth}
			\includegraphics[width=\linewidth,page=1]{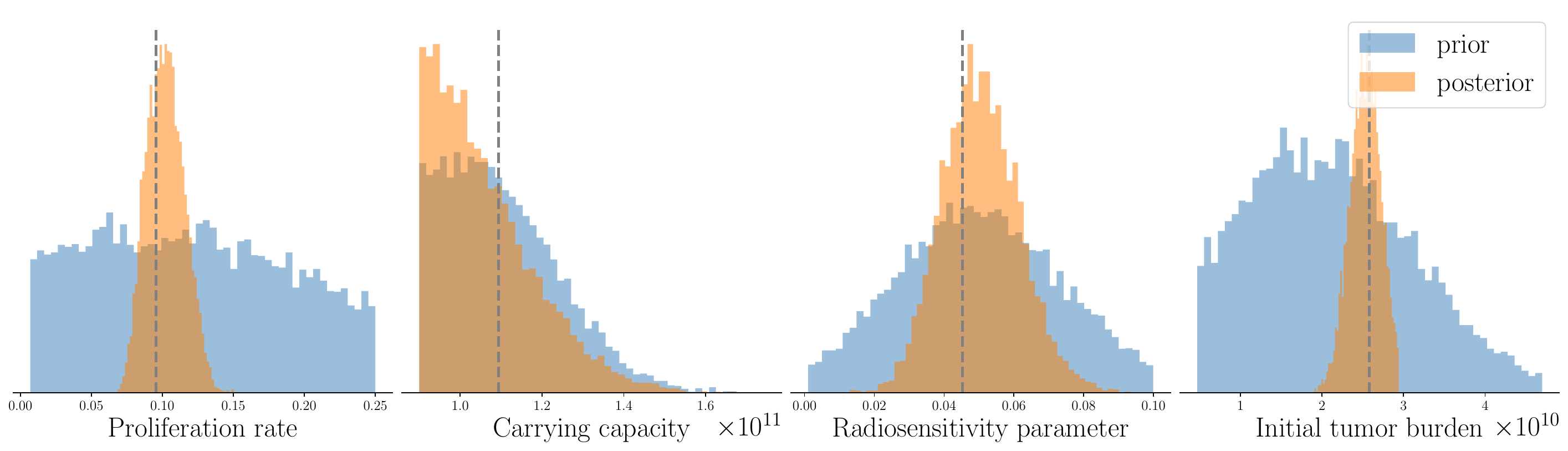}
			\caption{Patient 2.}
			\label{fig:p2inv_dist}
		\end{subfigure}
		\begin{subfigure}{0.9\textwidth}
			\includegraphics[width=\linewidth,page=1]{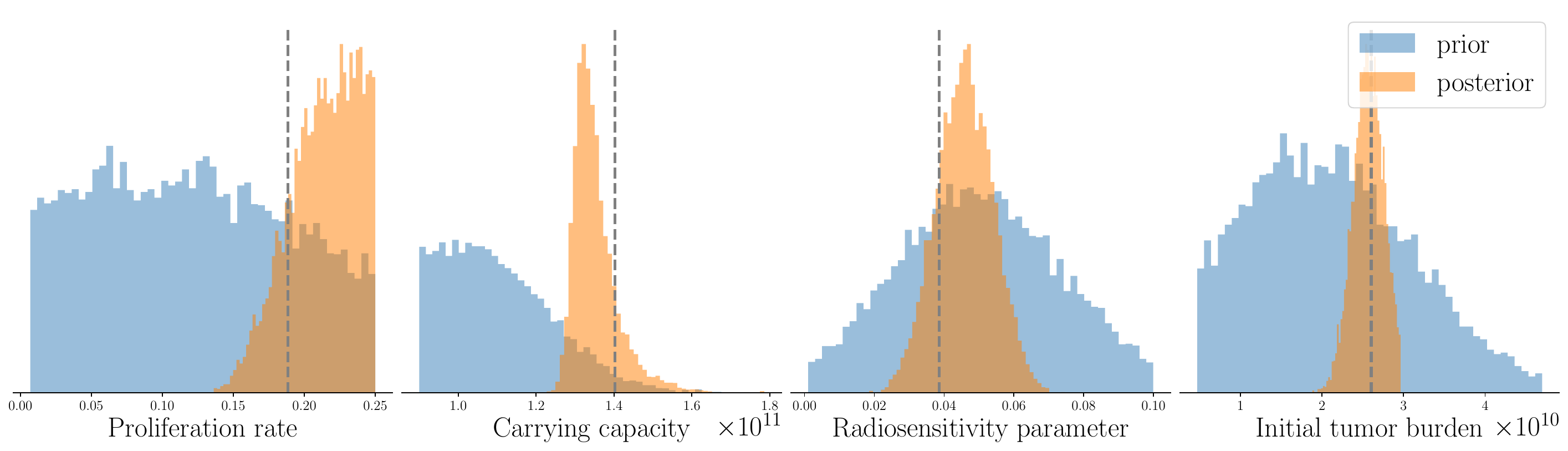}
			\caption{Patient 3.}
			\label{fig:p3inv_dist}
		\end{subfigure}
		
		\setcounter{subfigure}{-1}
		\caption{Posterior parameter distributions of the calibrated digital twins after assimilating observed data at the three imaging visits are shown for three patients: (a) Patient 1, (b) Patient 2, and (c) Patient 3. The dashed gray line indicates true parameter values for reference and not seen by the digital twin. Note that the same prior distribution is used for initializing digital twins of all the patients. Posterior distributions concentrate around the unseen true parameters and show reduction in uncertainty compared to prior distributions.}
		\label{fig:post_params}
	\end{figure}

	The patient-specific calibrated digital twin is obtained by assimilating the three observational data available for each patient as described in Section~\ref{s:inv}, while the priors of all patients are obtained from population clinical data as described in Table~\ref{tab:prior}. The histograms in Figure~\ref{fig:post_params} show the prior and the posterior distributions for the probabilistic model parameters $\theta$ in the digital state for the three case-study patients. We observe that the posterior distributions concentrate around the (unseen) true physical state $\theta_\text{true}$ given in Table~\ref{tab:case_params} (shown by dashed lines in Figure~\ref{fig:post_params}). We can also see that the overall uncertainty in the probabilistic parameters reduces compared to the prior distribution. In the clinical context, this could be interpreted as updating our belief about the specific patient's tumor dynamics as more data is collected during the clinical management of the disease. 
	
\begin{figure}[!htb]
		\centering
		\begin{subfigure}{0.9\textwidth}
			\includegraphics[width=\linewidth,page=2]{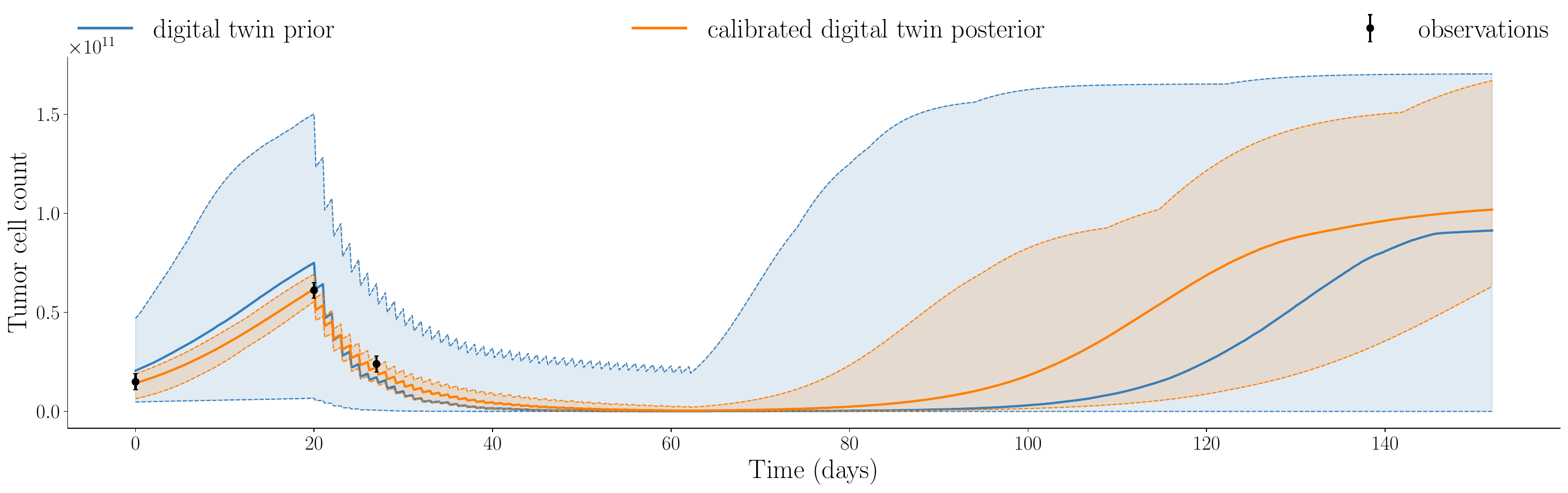}
			\caption{Patient 1.}
			\label{fig:p1inv_traj}
		\end{subfigure}
		\begin{subfigure}{0.9\textwidth}
			\includegraphics[width=\linewidth,page=2]{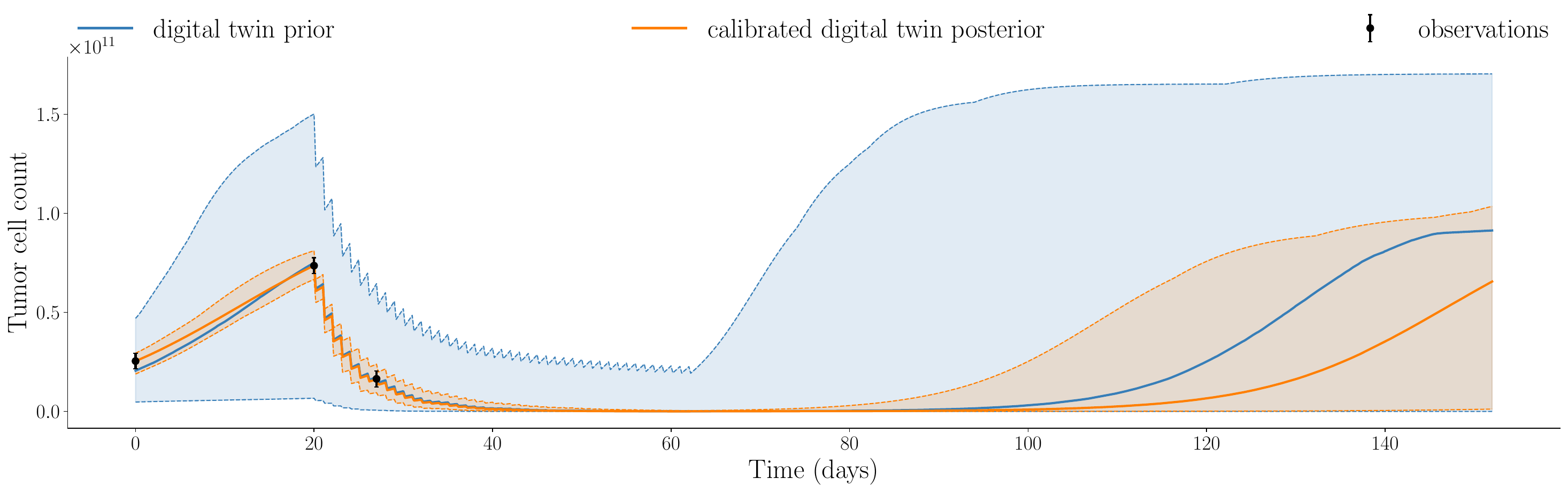}
			\caption{Patient 2.}
			\label{fig:p2inv_traj}
		\end{subfigure}
		\begin{subfigure}{0.9\textwidth}
			\includegraphics[width=\linewidth,page=2]{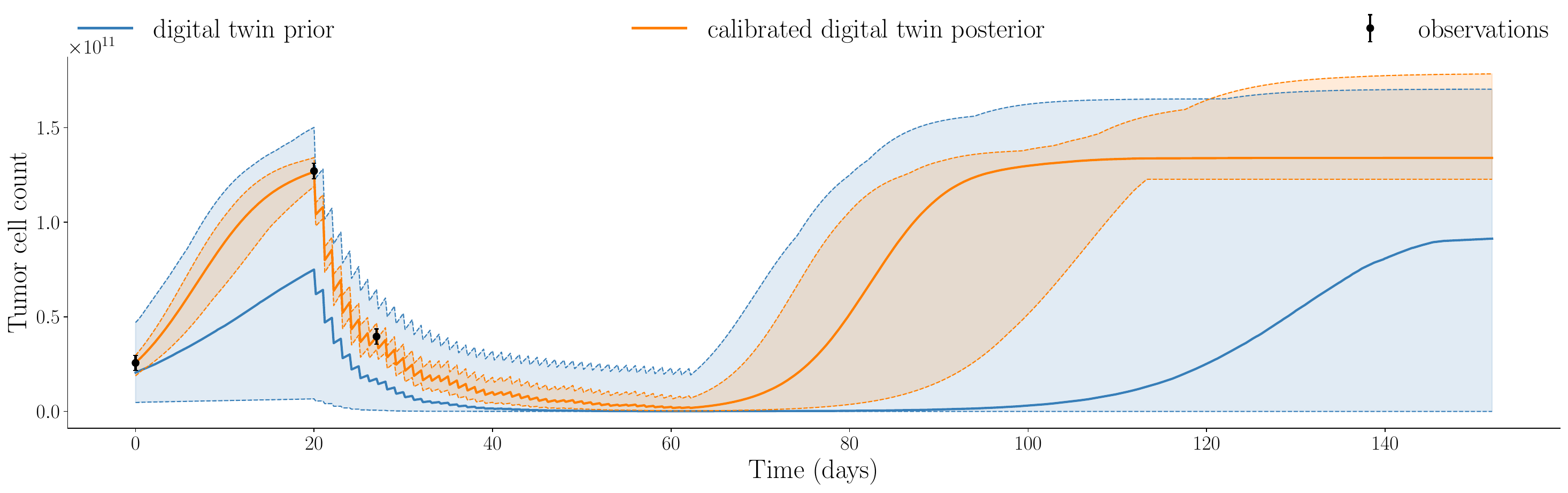}
			\caption{Patient 3.}
			\label{fig:p3inv_traj}
		\end{subfigure}
		\caption{Calibrated digital twin posterior trajectory after assimilating observed data at the three imaging visits compared to prior trajectory for the three patients: (a) Patient 1, (b) Patient 2, and (c) Patient 3 (median shown by solid line; minimum and maximum shown by dashed lines). Posterior trajectories capture the observed data and reduce uncertainty compared to prior trajectories.}
		\label{fig:post_traj}
	\end{figure}

	To forecast model uncertainty, the uncertainty in the probabilistic model parameters is propagated forward by sampling $10,000$ parameter sets from the joint posterior obtained by MCMC and solving the mechanistic model in Eq.~\eqref{e:forward_ode} to generate the posterior trajectories. Figure~\ref{fig:post_traj} shows the posterior trajectories for the three case-study patients with the maximum and minimum bands. The posterior trajectories simulated from the calibrated digital twins exhibit tighter bounds (as shown by the orange posterior vs blue prior shaded regions) compared to the prior trajectories, which indicate a decrease in overall uncertainty. This is a consequence of our updated belief about the patient's tumor dynamics resulting from the integration of longitudinal data into the digital twin for each individual patient. We also see that the observations for each patient are captured within the shaded region of the respective posterior trajectories, which indicates the range of posterior trajectories. The quantification of uncertainty in model outputs is crucial for estimating the risk of tumor propagation used for risk-aware clinical decision-making. 
	
	\subsection{Patient-specific treatment optimization under uncertainty} \label{s:res_ouu}
\begin{figure}[!htb]
		\centering
		\begin{subfigure}{0.32\textwidth}
			\includegraphics[width=\linewidth,page=3]{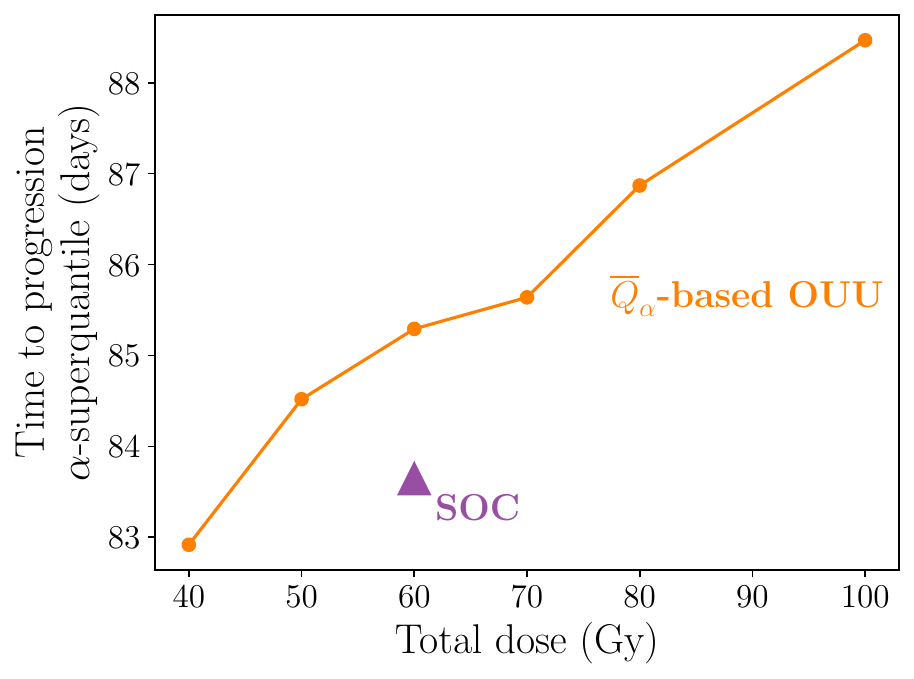}
			\caption{Patient 1.}
			\label{fig:p1pareto}
		\end{subfigure} \hfill
		\begin{subfigure}{0.32\textwidth}
			\includegraphics[width=\linewidth,page=3]{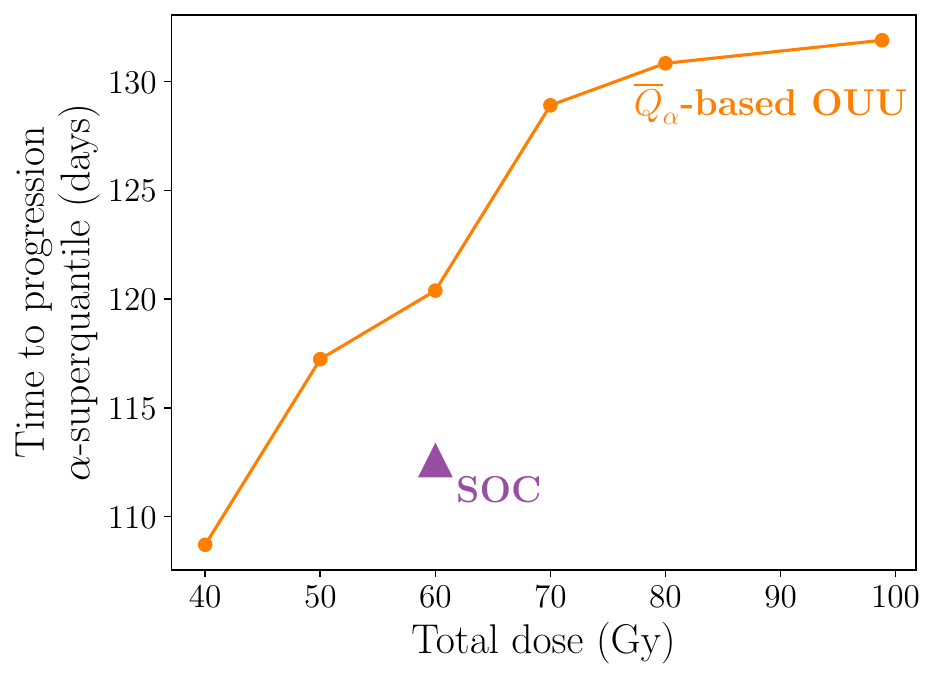}
			\caption{Patient 2.}
			\label{fig:p2pareto}
		\end{subfigure} \hfill
		\begin{subfigure}{0.32\textwidth}
			\includegraphics[width=\linewidth,page=3]{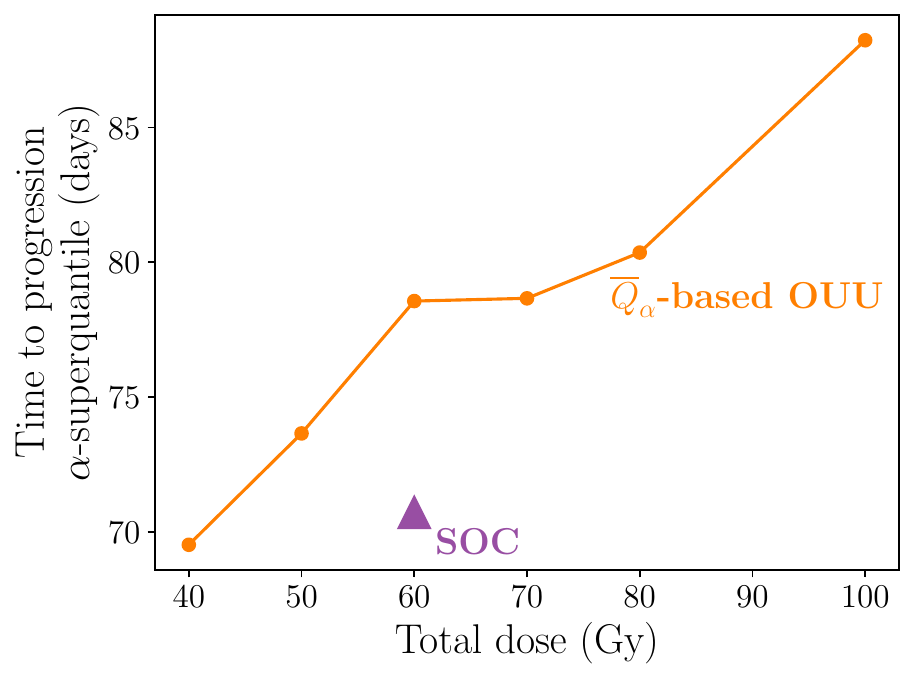}
			\caption{Patient 3.}
			\label{fig:p3pareto}
		\end{subfigure}
		\caption{Pareto front showing the suite of patient-specific optimal therapy solutions obtained from the multi-objective risk-based OUU compared to SOC treatment for the three patients: (a) Patient 1, (b) Patient 2, and (c) Patient 3. The treatment plans from OUU solutions show increased TTP compared to the SOC.}
		\label{fig:pareto}
	\end{figure}

	We now present the risk-aware optimal treatment regimens obtained using the calibrated digital twins. We use the $\alpha$-superquantile risk measure for quantifying risk of tumor growth as described in Section~\ref{s:risk} and then solve a risk-based multi-objective problem to obtain a suite of patient-specific optimal treatment regimens under uncertainty as described in Section~\ref{s:ouu}. Figure~\ref{fig:pareto} shows the results of our patient-specific treatment optimization for the three case-study patients. We plot the TTP $\alpha$-superquantile against the dose threshold to show the Pareto front indicating the trade-off between the two quantities. Figure~\ref{fig:pareto} shows that the SOC regimen is not the optimal dosing schedule in multiple ways. First, for the applied radiation dose of 60 Gy, the optimal radiotherapy plan can demonstrate superior tumor control in terms of longer TTP. Second, similar or better tumor control can be exerted with a reduced amount of dose compared to SOC, as shown by the TTP values at the $40$ Gy and $50$ Gy levels. Third, greater tumor control than the SOC regimen can be exerted for total doses $\ge 60$ Gy as seen in Figure~\ref{fig:pareto}. 
	
\begin{figure}[!htb]
		\centering
		\begin{subfigure}{0.9\textwidth}
			\includegraphics[width=\linewidth,page=4]{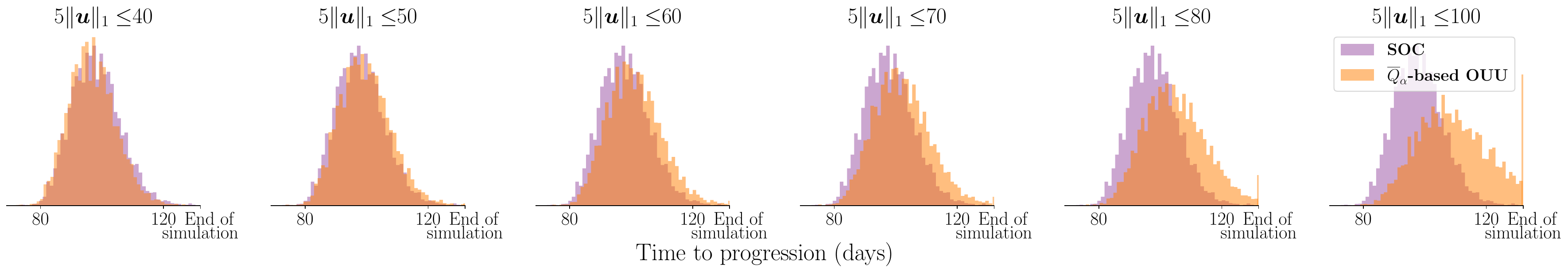}
			\caption{Patient 1.}
			\label{fig:p1ttp_dist}
		\end{subfigure}\hfill
		\begin{subfigure}{0.9\textwidth}
			\includegraphics[width=\linewidth,page=4]{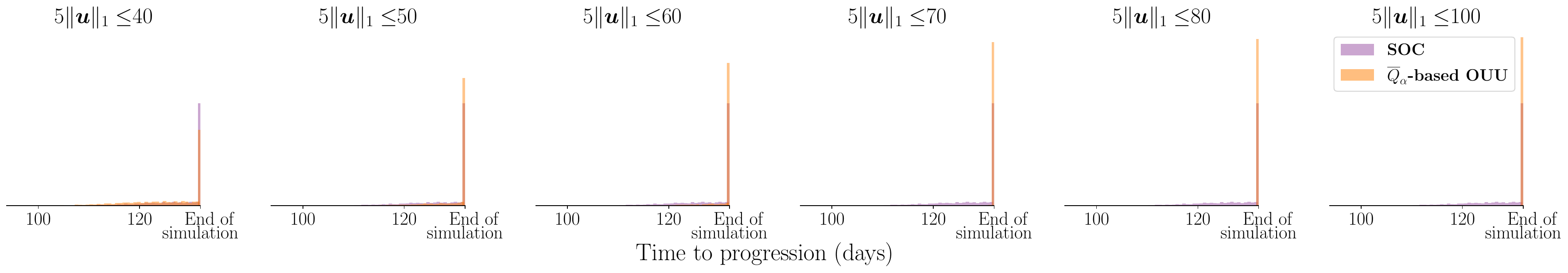}
			\caption{Patient 2.}
			\label{fig:p2ttp_dist}
		\end{subfigure} \hfill
		\begin{subfigure}{0.9\textwidth}
			\includegraphics[width=\linewidth,page=4]{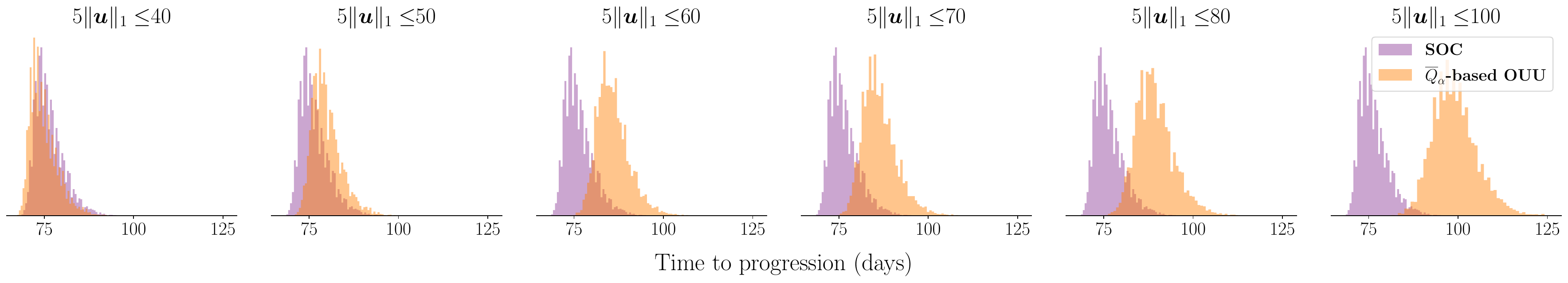}
			\caption{Patient 3.}
			\label{fig:p3ttp_dist}
		\end{subfigure}
		\caption{Comparing TTP distribution using optimized treatment under uncertainty vs SOC treatment for the three case-study patients: (a) Patient 1, (b) Patient 2, and (c) Patient 3. Note that ``end of simulation'' refers to the maximum TTP value of 132 days stipulated by $t_\text{finite}=152$ days. We can see the optimized treatment plans lead to separation in TTP distributions compared to the SOC and move towards higher TTP values indicating superior tumor control.}
		\label{fig:ttp_dist}
	\end{figure}

	Figure~\ref{fig:ttp_dist} shows the entire distribution of the TTP using the optimized treatment regimens compared to the SOC treatment for the three case-study patients.  For Patient 1 in Figure~\ref{fig:p1ttp_dist}, we observe that with total dose constraint $D_\text{max}\ge 60$ Gy we see a progressive separation between the TTP distributions obtained from the SOC and the optimized treatment plans as we increase the total dose. The TTP distribution for optimized plans move towards higher TTP values. Thus, the optimal treatment plans lead to TTP distributions that are therapeutically superior than the ones obtained from the SOC. Additionally, considering total doses lower than the 60 Gy administered by the SOC, we observe that there is not much separation in the distributions of TTP. This result highlights the possibility of reducing the total dose administered in the SOC protocol by 10-20 Gy while obtaining a similar tumor control for Patient 1. A similar trend in separation of TTP distributions is seen for Patient 3, as shown in Figure~\ref{fig:p3ttp_dist}. In fact, the separation is obvious even at a total dose constraint of 50 Gy, for which the TTP distribution obtained from the risk-aware optimization already indicates superior therapeutic outcomes than the one obtained from the SOC. At higher doses, the TTP distributions from the optimized results shift towards higher TTP values for patient 3 and shows separation from the TTP distribution obtained from the SOC. For Patient 2 in Figure~\ref{fig:p2ttp_dist}, the most likely outcome is that the disease is controlled within the finite time horizon considered in this study (i.e., $t_\text{finite}=152$ days). However, the optimized treatment plans still increase the probability of treatment success. This can be seen by the higher peaks at the ``end of simulation'' in  Figure~\ref{fig:p2ttp_dist}.
	
\begin{figure}[!htb]
		\centering
		\begin{subfigure}{0.32\textwidth}
			\includegraphics[width=\linewidth,page=5]{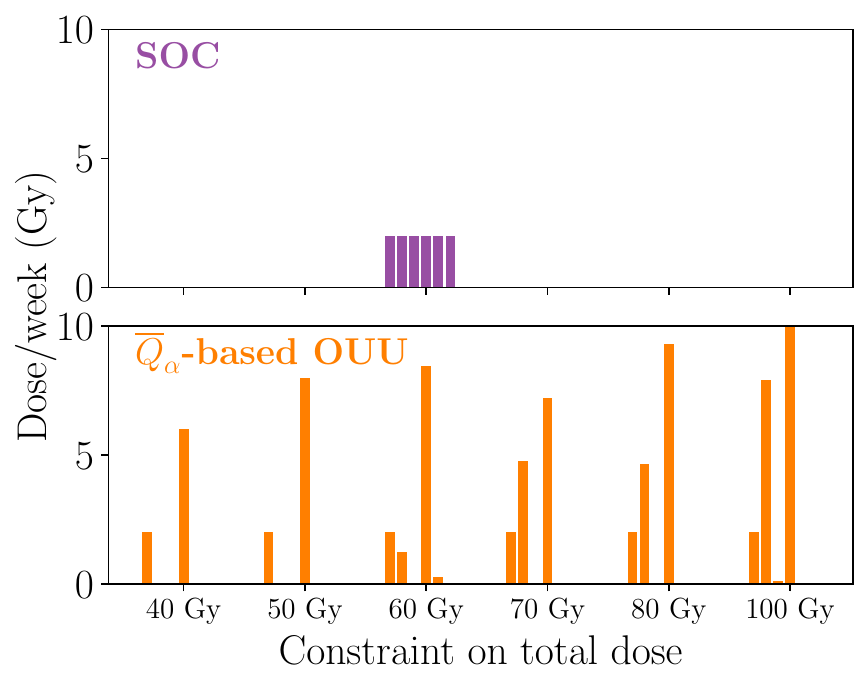}
			\caption{Patient 1.}
			\label{fig:p1opt_dose}
		\end{subfigure} \hfill
		\begin{subfigure}{0.32\textwidth}
			\includegraphics[width=\linewidth,page=5]{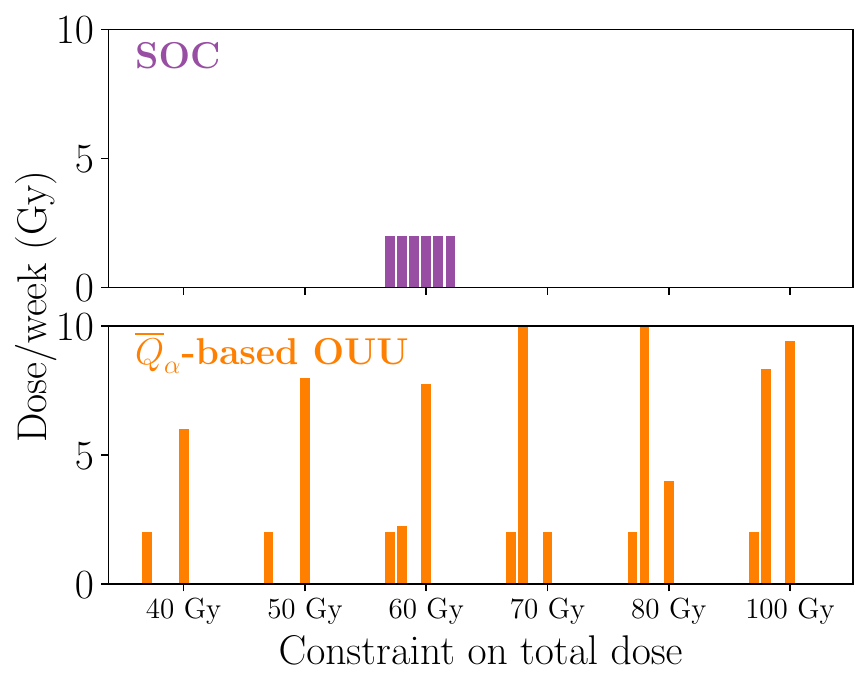}
			\caption{Patient 2.}
			\label{fig:p2opt_dose}
		\end{subfigure} \hfill
		\begin{subfigure}{0.32\textwidth}
			\includegraphics[width=\linewidth,page=5]{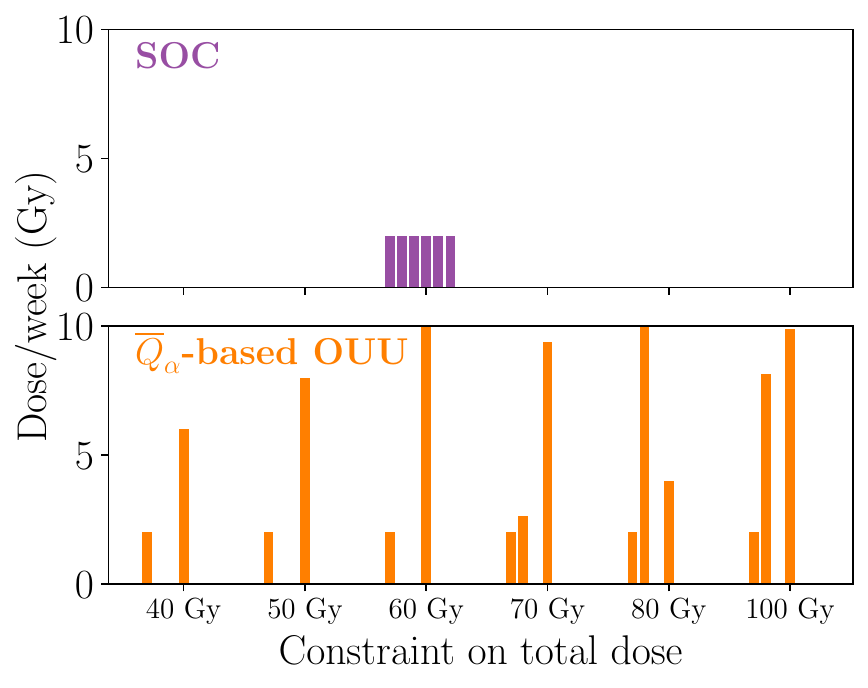}
			\caption{Patient 3.}
			\label{fig:p3opt_dose}
		\end{subfigure}
		\caption{Optimized dose schedules reveal different strategies on a patient-specific level as well as for the solutions along the Pareto front for the three patients: (a) Patient 1, (b) Patient 2, and (c) Patient 3. Generally, optimized dose schedules result in larger doses towards the end of the treatment time frame.}
		\label{fig:opt_dose}
	\end{figure}

	The SOC and optimized treatment regimens for the three case-study patients are visualized in Figure~\ref{fig:opt_dose}, which shows the dose schedule over the six weeks of RT. At lower total dose levels, treatment plans tend to feature one large dose roughly halfway through the six week treatment course. At intermediate total doses, there is a secondary dose similar to the SOC in the second week of treatment and a large dose applied towards the end of the treatment timeframe. Finally, strategies featuring a higher total dose tend to feature one large dose in the second week of treatment and then another dose towards the end of the therapy. This second dose can be larger or lower than the one delivered the second week of treatment, but it tends to be larger than the corresponding SOC dose. Overall, Figure~\ref{fig:opt_dose} shows that optimized plans tend to leave weeks without RT with only the chemotherapy being administered during that time. This can be viewed as an additional advantage with not requiring patients to come in every week.

\begin{figure}[!htb]
		\centering
		\includegraphics[width=0.7\textwidth, page=1]{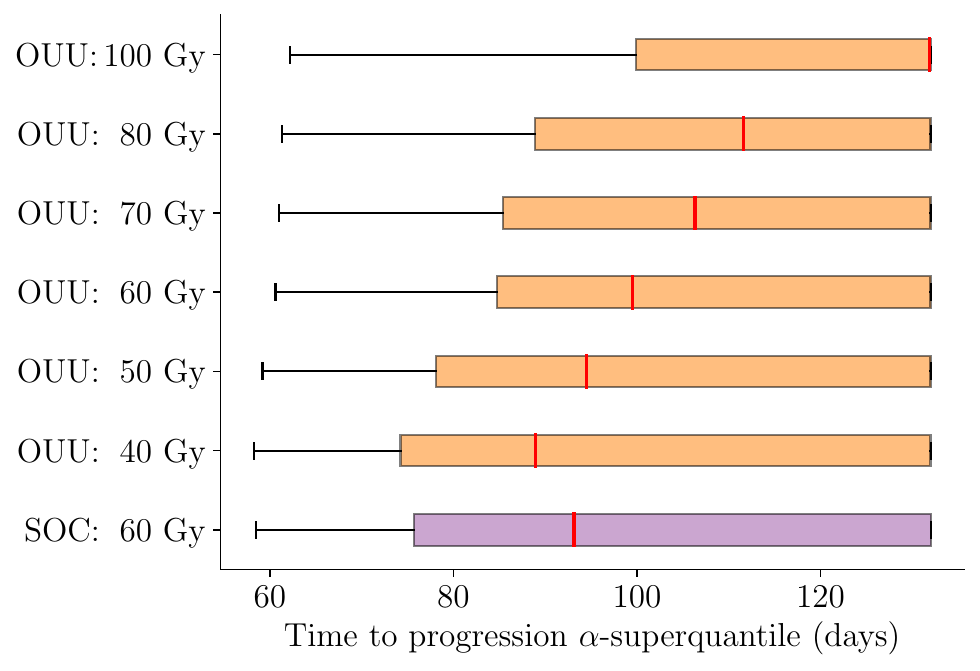}
		\caption{Comparison between optimal treatment regimens under uncertainty and SOC for TTP $\alpha$-superquantile over the cohort of 100 patients. The solution of the optimization problem in Eq.~\eqref{e:multiobj_formulation} with $D_\text{max}=$ X Gy is indicated as ``OUU: X Gy''. Optimal treatment regimens lead to better TTP with $D_\text{max}\ge 60$ Gy and similar TTP with $D_\text{max}< 60$ Gy when compared to the SOC}
		\label{fig:ouu_100}
	\end{figure}

\begin{figure}[!htb]
		\centering
		\begin{subfigure}{0.49\textwidth}
			\includegraphics[width=\linewidth,page=2]{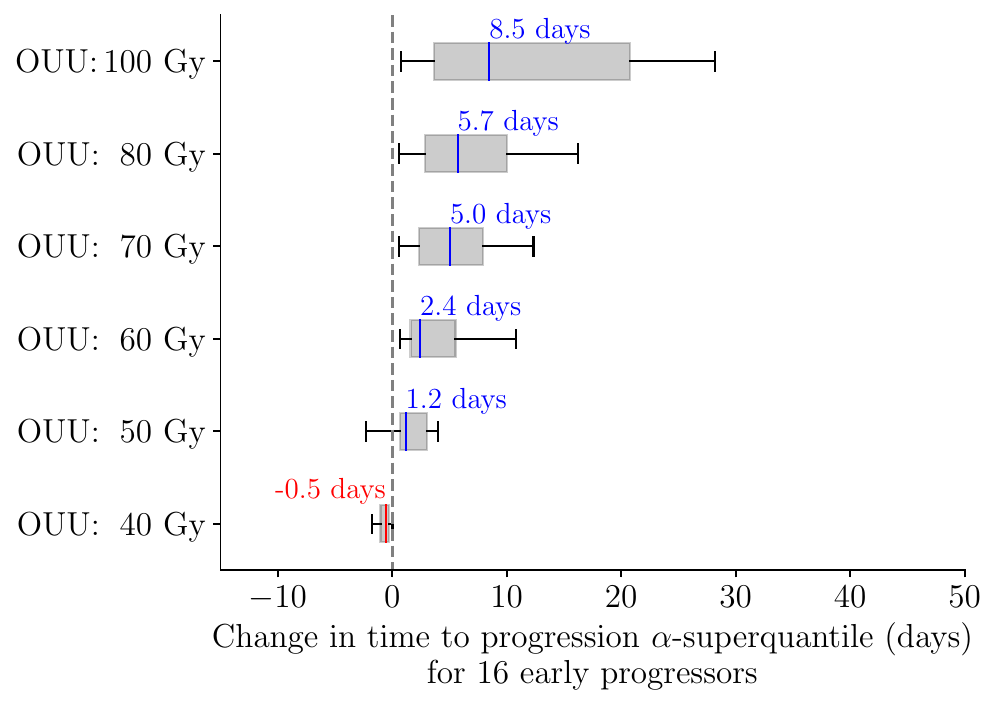}
			\caption{}
			\label{fig:change_ttp_early}
		\end{subfigure} \hfill
		\begin{subfigure}{0.49\textwidth}
			\includegraphics[width=\linewidth,page=4]{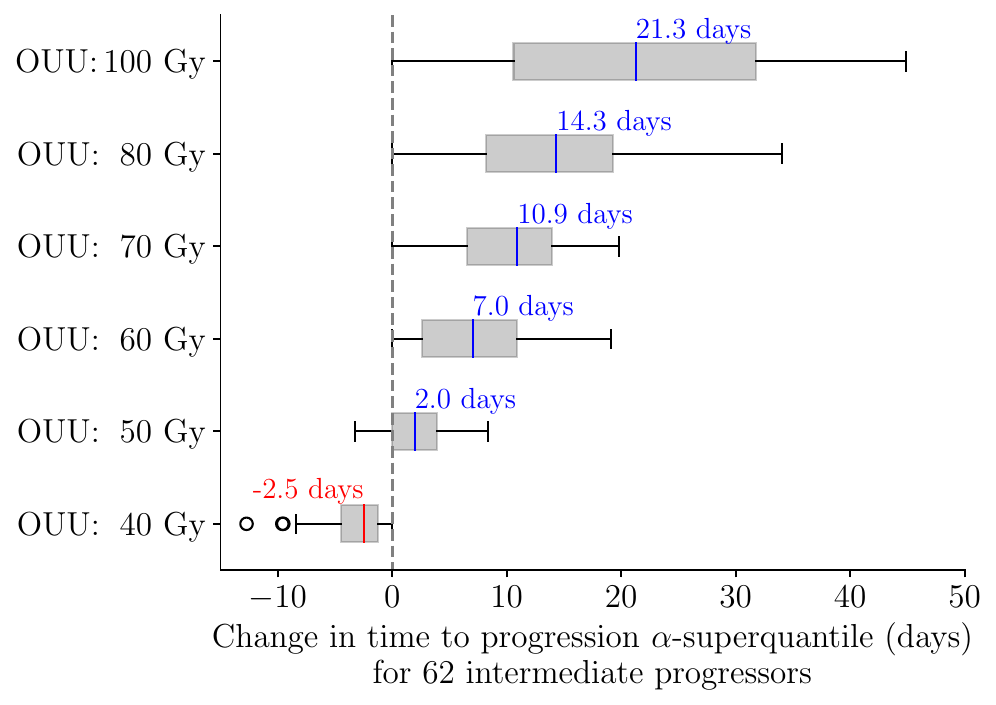}
			\caption{}
			\label{fig:change_ttp_inter}
		\end{subfigure}
		\caption{Change in TTP $\alpha$-superquantile when using optimal treatment regimens as compared to SOC for (a) early progressors: 16 patients with TTP $\alpha$-superquantile less than 1 month after end of RT, and (b) intermediate progressors: 62 patients  with TTP $\alpha$-superquantile between 1-3 months after end of RT. Median values are indicated in text above each box plot. The solution of the optimization problem in Eq.~\eqref{e:multiobj_formulation} with $D_\text{max}=$ X Gy is indicated as ``OUU: X Gy''. The median change indicates longer TTP compared to the SOC when optimal treatments are used.}
		\label{fig:change_ttp}
	\end{figure}

	We now analyze the performance of the predictive digital twins at the cohort level relative to the SOC and highlight the performance for the three patient groups defined in Section~\ref{s:study}. To this end, we deploy the predictive digital twins for each of the 100 \textit{in silico} patients to obtain patient-specific optimal RT treatment regimens. Figure~\ref{fig:ouu_100} shows the boxplot of TTP $\alpha$-superquantile values obtained for the patient-specific optimal treatment regimens and the SOC treatment. We denote the solution of the optimization problem in Eq.~\eqref{e:multiobj_formulation} with $D_\text{max}=$ X Gy as ``OUU: X Gy'' in Figure~\ref{fig:ouu_100}. We can see that the optimal treatment regimens lead to better TTP compared to the SOC over the cohort of 100 patients with $D_\text{max}\ge 60$ Gy. 
	We further analyze the change in TTP $\alpha$-superquantile given by ($\text{(TTP $\alpha$-superquantile from ``OUU: X Gy'')}-\text{(TTP $\alpha$-superquantile from SOC)}$) for each patient in the different patient groups. Figure~\ref{fig:change_ttp} shows the boxplots for the change in TTP $\alpha$-superquantile values for early and intermediate progressors. Figure~\ref{fig:change_ttp_early} captures the response to optimal treatment for early progressors. For optimal solutions ``OUU: 60 Gy'', which use the same total dose as the SOC, we obtain a median change in TTP $\alpha$-superquantile of $+2.4$ days with all 16 early progressors having better TTP $\alpha$-superquantile values compared to the SOC. The higher dose optimal treatment regimens provide other therapeutic options to the patient and the treating physician with up to $+8.5$ days median change in TTP $\alpha$-superquantile.
	Figure~\ref{fig:change_ttp_inter} captures the response to optimal treatment for intermediate progressors. In this case, the optimal solution ``OUU: 60 Gy'' leads to a median change in TTP $\alpha$-superquantile of $+7$ days with all 62 intermediate progressors having better TTP $\alpha$-superquantile values compared to the SOC. The higher dose optimal treatment regimens provide additional therapeutic strategies with up to $+21.3$ days median change in TTP $\alpha$-superquantile.
	
\begin{figure}[!htb]
		\centering
		\includegraphics[width=0.9\textwidth, page=6]{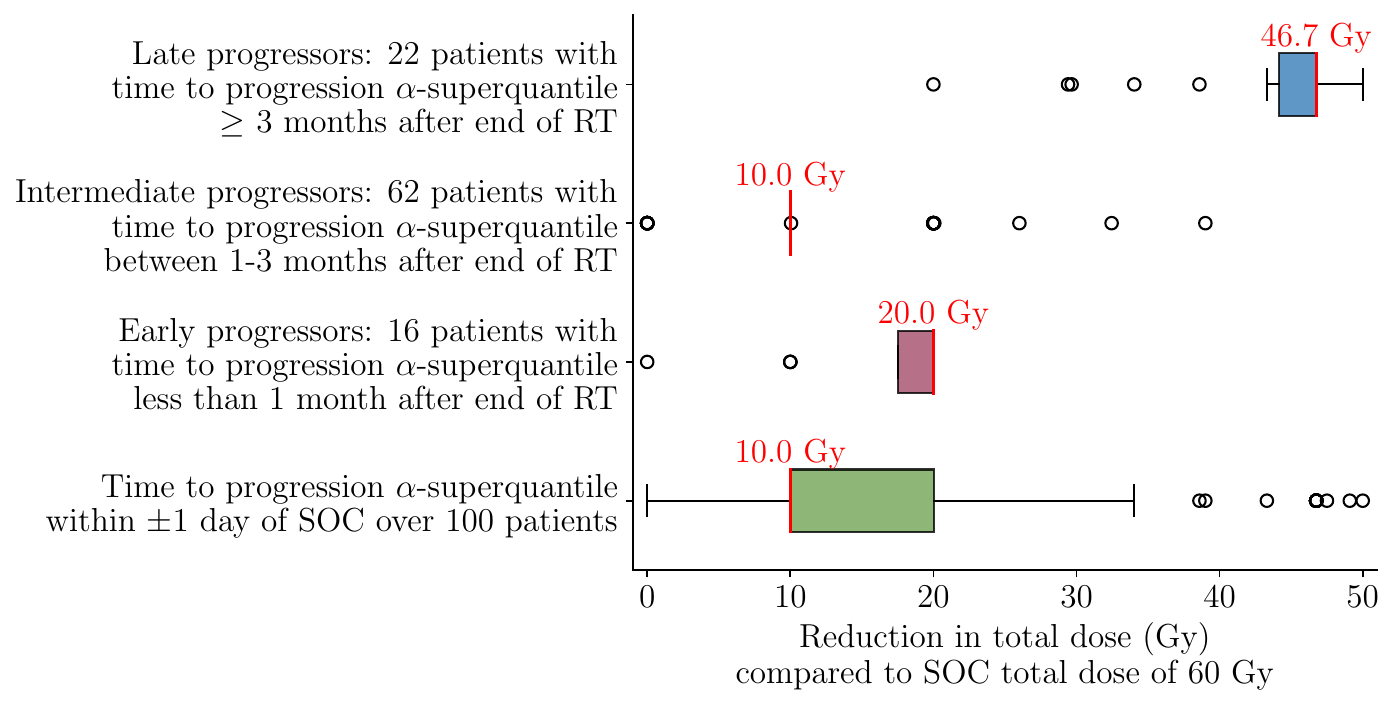}
		\caption{Reduction in total dose compared to SOC total dose of 60 Gy to achieve TTP $\alpha$-superquantile within $\pm 1$ day of SOC over the cohort of 100 patients and for the different patient groups of early, intermediate, and late progressors. Median values are indicated in red text above each box plot. Optimal treatment plans lead to 16.7$\%$ median dose reduction over the 100 patient cohort and $77.8\%$ median dose reduction for the late progressors compared to the SOC.}
		\label{fig:red_dose_late}
	\end{figure}

	The late progressors are not expected to have any reduction in TTP since almost all treatment options reach the TTP $\alpha$-superquantile maximum value of 132 days. One can see that the definition of late progressors coincides with the finite time horizon considered. Instead, the late progressors can have similar tumor control with reduction in total required dose. Recall that to achieve this therapeutic objective of mitigating the toxicity effects with our digital twin formulation, we use the penalty parameter $\lambda=0.001$ in Eq.~\eqref{e:multiobj_formulation} to drive the optimizer to lowest possible dose while obtaining the same level of tumor control, as described in Section~\ref{s:ouu}. Figure~\ref{fig:red_dose_late} shows that this approach achieves a reduction in total radiation dose while maintaining a similar treatment efficacy as SOC for the entire cohort of 100 patients as well as the three patient groups. We can see that risk-aware optimal treatment plans can lead to a median of 10 Gy (16.7$\%$) reduction in total dose compared to the SOC over all 100 patients, while keeping the TTP $\alpha$-superquantile within $\pm 1$ day of that obtained for the SOC. Analyzing the effect on the different patient groups provides additional insight on the possible amount of dose reduction depending on the underlying  dynamics of tumor response. For the early and intermediate progressors, we can see a median reduction in dose of 10 Gy and 20 Gy, respectively. However, for the late progressors we can see that the median reduction in dose is $46.7$ Gy. This comparatively higher decrease in dose is a direct consequence of slower tumor growth and lower number of tumor cells in the tumors of late progressors within the post-treatment timeframe considered in this study.
	
	\subsection{Survival analysis of optimal treatments}\label{s:res_survival}
\begin{figure}[!htb]
		\centering
		\resizebox{\textwidth}{!}{%
			\input{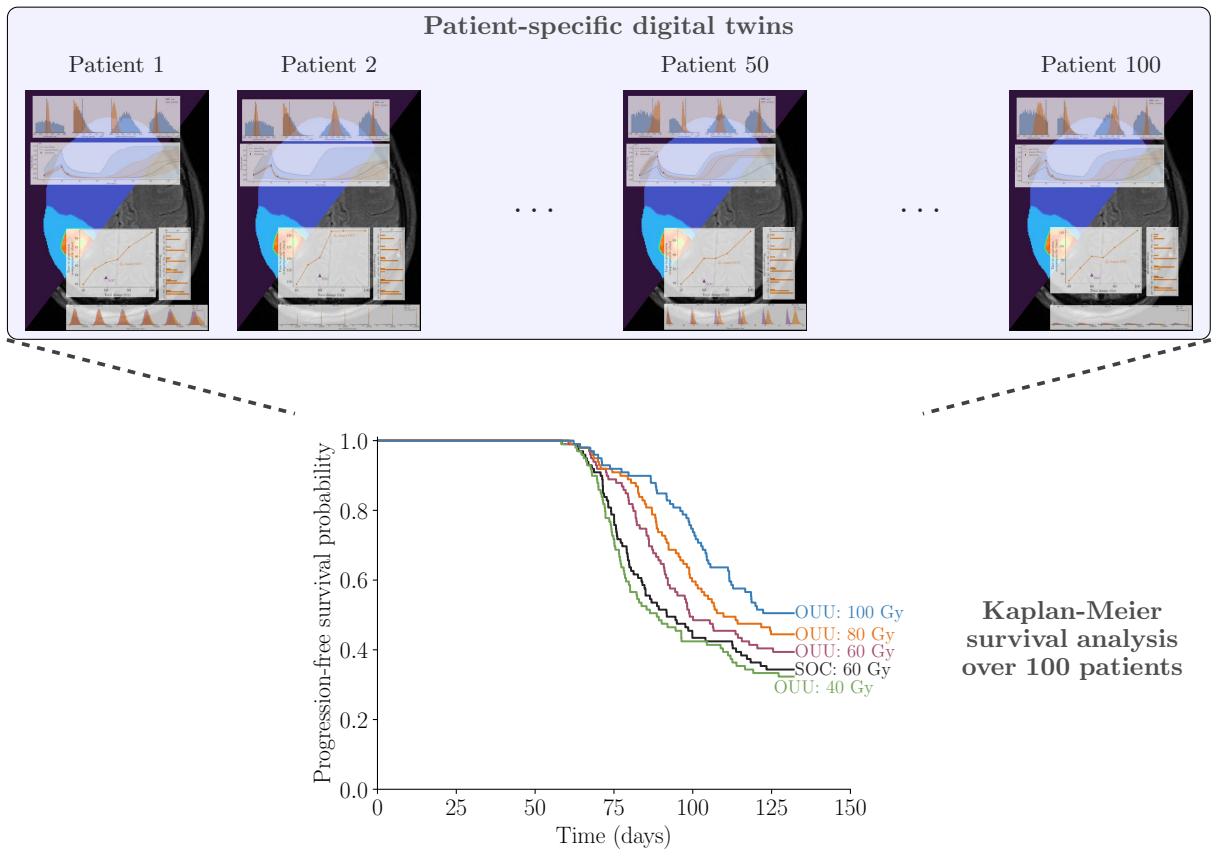}
		}
		\caption{Illustrating patient-specific digital twins for cohort of 100 patients and analyzing the performance of proposed optimal treatment plans vs SOC through Kaplan-Meier survival analysis using the TTP $\alpha$-superquantile values.}
		\label{fig:km}
	\end{figure}

	We now show the results for the Kaplan-Meier survival analysis described in Section~\ref{s:km}. The results on survival analysis indicate that making optimal decisions at the individual level using patient-specific predictive digital twins provide either the same survivability as SOC plans or improve it across the cohort of patients. The survival curves in Figure~\ref{fig:km} show that all optimal therapeutic regimens generated with a total dose threshold greater than or equal to that of the SOC plan of 60 Gy have survival curves that are superior to the corresponding curve obtained with SOC plan. There is a clear gap between the 80 Gy and 100 Gy plans and the SOC with logrank p-values of 0.005 and 0.0002, respectively. For the optimized therapy with maximum allowable dose of $60$ Gy, the survival curve is statistically superior to the SOC treatment option with p-value of 0.05. Figure~\ref{fig:km} further shows that the 40 Gy plans show  almost no difference in tumor control when compared to the SOC. With a p-value of $0.35 \gg 0.05$, we conclude that the optimized treatment plans with maximum allowable dose of 40 Gy is not significantly different than the SOC, which uses 60 Gy. 

	\section{Discussion} \label{s:discuss}
	Our digital twin methodology provides the computational and mathematical foundation to dynamically integrate patient data with any given model of tumor growth and response. The predictive digital twin can provide risk-aware personalized treatment plans and we demonstrate it through optimized RT treatment regimens for HGG. There is a rich literature of modeling glioma growth and response across different scales \cite{Alfonso2017,hormuth_addr_2022}; however, there are limited approaches that integrate these models with patient-specific data to optimize or adapt therapy in a dynamic fashion while accounting for underlying uncertainty. For development and demonstration of this framework, we employed an ODE model describing the total tumor cell proliferation. Response to radiotherapy and chemotherapy were reduced to an instantaneous effect at the time of treatment. While the model’s biological detail is sufficient to demonstrate from end-to-end the key components of the predictive digital twin, a more biologically complex model may be needed in the clinical setting to account for inter-tumor heterogeneity in treatment delivery, proliferation rates, and resistance to therapy. Brain tumor growth and response to treatment has been described via a more complex reaction-diffusion model \cite{Alfonso2017,Hormuth21SciRep,Swanson2000} that describes the spatial-temporal evolution of tumors cells throughout the brain due to invasion (i.e., the diffusion term) and proliferation (i.e., the reaction term). Our predictive digital twin can be readily adapted to other more complex models and cancer types where the prerequisite spatial-temporal or temporal data are available. We plan to extend the predictive digital twin to a more descriptive partial differential equation model that can extract the spatial information, but comes with a much higher computational cost challenge. While this digital twin focused on the delivery of RT concurrently with chemotherapy, our predictive digital twin could be adapted to other treatment modalities (e.g., immunotherapy, tumor treatment fields, convection-enhanced delivery) provided the pre-requisite data and models are available. We will also explore the critical issue of when to get the patient in to gather more imaging data.
	
	We stress that the overall motivation of this work was the development of a digital twin framework which could be applied to any biology-based model of tumor growth. However, clinical treatment should ultimately be based upon established clinical guidelines~\cite{NCCN2022}. Indeed, one limitation of this study is the use of an \textit{in silico} patient cohort whose growth and response parameters are sampled from literature values and governed by our model. We would anticipate that when applied to actual patient data, the precise gains between different RT regimen would differ to those presented in this work due to variability in the patient population and limitations of our current model. Therefore, a clinical trial would be necessary to establish the benefit of digital twin adapted RT regimens.
	
	The optimized treatment regimens identified here tended towards a more hypofractionated (i.e., larger dose per day over a shorter period of time) or intermittent paradigm for treatment. While not widely used in the standard-of-care setting for initial treatment, hypofractionated radiotherapy is safe and tolerable \cite{Gutin2009} and employed in the recurrent setting and for patients with poor prognoses \cite{Trone2020}. Compared to standard treatment schedules, hypofractionated radiotherapy has an increased cell kill over a shorter period of time, potential immunogenic effects, but at a potentially increased neuro-toxicity \cite{Reznik2018,Hingorani2012}. In the context of first-line treatment following surgery for HGG, the benefits of hypofractionated radiotherapy are less clear but some studies have reported similar tumor control and a palliative benefit due to its condensed treatment schedule \cite{Floyd2004}. Other modeling approaches from \cite{Randles2021,Leder2014, Brueningk2021}have also identified optimized therapies that do not conform to the standard treatment paradigm. Recently, in the pre-clinical setting, \cite{Randles2021} predicted improved outcomes with a hyperfractioned regimen (i.e., smaller doses more than once a day) compared to standard dosing and this was validated via experiments. At the clinical level, \cite{Brueningk2021} explored intermittent radiation therapy (i.e., 6 Gy $\times$ 1 day every 6 weeks) in comparison to hyperfractionated radiotherapy for recurring disease. \cite{Brueningk2021} observed similar tumor control for both scenarios, but prolonged time to progression when additional intermittent doses of radiotherapy were delivered.  In a future work, we plan to explore different clinical objectives and treatment controls, which could lead to other types of treatment regimens with further improvement in treatment efficacy.

	\section{Conclusions} \label{s:conclusion}
	We have developed a predictive digital twin methodology that enables end-to-end uncertainty quantification and optimization of personalized treatment regimens under uncertainty. The predictive digital twins employ a Bayesian perspective to account for uncertainty through the entirety of the clinical process to capture our knowledge (or lack thereof) of the individual dynamics. We illustrate the effectiveness of the predictive digital twin to optimize RT treatment plans for individual high-grade glioma patients. This subclass of tumors are aggressive and highly heterogeneous in nature, motivating the need for patient-specific modeling and treatment planning. We applied the predictive digital twin to a cohort of \textit{in silico} patients with response and growth characteristics assigned from literature studies. We calibrated the digital twin with patient-specific data and used the calibrated digital twin to inform risk-aware clinical decision-making. We solve a multi-objective risk-based optimization under uncertainty problem to provide a suite of treatment plans balancing tumor control and toxicity. The risk-aware patient-specific optimization of the dose for the \textit{in silico} study demonstrated that we were able to extend the progression-free survival for the standard total dose of 60 Gy relative to the standard-of-care. Additionally, we were able to identify potential reductions in the total dose delivered for similar tumor control response as the standard-of-care. These results represent a first step towards a practical digital twin for treatment optimization and future studies should expand this approach to data collected in the clinical setting.

	\section*{Author Contributions}
	All authors contributed to conception and design of the study. AC, GP, MK, and KW contributed to the computational methodology. AC generated the results and plots. AC, GP, and DAH wrote the first draft of manuscript. All authors contributed to manuscript revision, read, and approved the submitted version.
	
	\section*{Funding}
	A.C. and K.W. acknowledge support from DARPA grant number DE-AC05-76RL01830 under the Automating Scientific Knowledge Extraction and Modeling (ASKEM) program and Department of Energy (DOE) Grant DE-SC0021239. K.W. acknowledges support from AFOSR MURI grant FA9550-21-1-0084. G.P. acknowledges support from the U.S. Department of Energy Computational Science Graduate Fellowship under Award Number DE-SC0021110. G.L. acknowledges the European Union’s Horizon 2020 research and innovation program under the Marie Sk\l{}odowska-Curie grant agreement No. 838786. T.E.Y. acknowledges support from the National Cancer Institute R01CA235800, U24CA226110, U01CA174706, and CPRIT RR160005. T.E.Y. is a CPRIT Scholar in Cancer Research. D.A.H. acknowledges support from CPRIT RP220225.
	
	This report was prepared as an account of work sponsored by an agency of the United States Government. Neither the United States Government nor any agency thereof, nor any of their employees, makes any warranty, express or implied, or assumes any legal liability or responsibility for the accuracy, completeness, or usefulness of any information, apparatus, product, or process disclosed, or represents that its use would not infringe privately owned rights. Reference herein to any specific commercial product, process, or service by trade name, trademark, manufacturer, or otherwise does not necessarily constitute or imply its endorsement, recommendation, or favoring by the United States Government or any agency thereof. The views and opinions of authors expressed herein do not necessarily state or reflect those of the United States Government or any agency thereof.

	\bibliographystyle{siam}
	\bibliography{dtoncobib}
	
\end{document}